\documentclass[amsmath,nofootinbib,toc,amssymb,pra,aps,showpacs,superscriptaddress,twocolumn,10pt]{revtex4-1}
\usepackage{amsmath,amsfonts,amssymb,amsthm,graphics,graphicx,epsfig,bbm}
\usepackage[colorlinks=true,citecolor=blue,linkcolor=red,urlcolor=blue]{hyperref}
\usepackage[usenames]{color}
\usepackage[english]{babel} 
\usepackage{graphicx}
\usepackage{lipsum, babel}
\usepackage{subfigure}
\usepackage{amsmath}
\usepackage{epsfig}
\usepackage{dcolumn}
\usepackage{bm}
\usepackage{color}
\usepackage{epstopdf}
\usepackage{amssymb}
\usepackage{amstext}
\usepackage{latexsym}
\usepackage{hyperref}
\usepackage{amsfonts}
\usepackage{psfrag}
\usepackage{xcolor}
\usepackage[normalem]{ulem}
\usepackage{dsfont}
\usepackage{txfonts}
\usepackage{overpic}
\usepackage{svg}
\usepackage{lettrine}
\renewcommand{\Re}{\mathrm{Re}}
\renewcommand{\Im}{\mathrm{Im}}
\newcommand{\ket}[1]{\vert #1 \rangle}
\newcommand{\bra}[1]{\langle #1 \vert}

\usepackage{setspace}
\newcommand{\red}[1]{\textcolor{red}{#1}}

\makeatletter
\def\l@section#1#2{}
\def\l@subsection#1#2{}
\makeatother
\begin{document}

\title{
\LARGE Simulating complex quantum networks with time crystals}

 \author{M. P. Estarellas}
     \thanks{These authors contributed equally.}
   \affiliation{National Institute of Informatics, 2-1-2 Hitotsubashi, Chiyoda-ku, Tokyo 101-8430, Japan}
 \author{T. Osada}
    \thanks{These authors contributed equally.}
  \affiliation{Tokyo University of Science, 1-3 Kagurazaka, Shinjuku, Tokyo, 162-8601, Japan}
    \affiliation{National Institute of Informatics, 2-1-2 Hitotsubashi, Chiyoda-ku, Tokyo 101-8430, Japan}
 \author{V. M. Bastidas}
 \thanks{These authors contributed equally.}
  \affiliation{NTT Basic Research Laboratories \& Research Center for Theoretical Quantum Physics,  3-1 Morinosato-Wakamiya, Atsugi, Kanagawa, 243-0198, Japan} 
 \author{B. Renoust}
 \affiliation{Osaka University, Institute for Datability Science, 2-8 Yamadaoka, Suita, Osaka Prefecture 565-0871, Japan} 
 \affiliation{National Institute of Informatics, 2-1-2 Hitotsubashi, Chiyoda-ku, Tokyo 101-8430, Japan}
 \affiliation{Japanese-French Laboratory for Informatics, CNRS UMI 3527, 2-1-2 Hitotsubashi, Chiyoda-ku, Tokyo 101-8430, Japan}
 \author{K. Sanaka}
  \affiliation{Tokyo University of Science, 1-3 Kagurazaka, Shinjuku, Tokyo, 162-8601, Japan}
 \author{W. J. Munro}
 \affiliation{NTT Basic Research Laboratories \& Research Center for Theoretical Quantum Physics,  3-1 Morinosato-Wakamiya, Atsugi, Kanagawa, 243-0198, Japan} 
 \affiliation{National Institute of Informatics, 2-1-2 Hitotsubashi, Chiyoda-ku, Tokyo 101-8430, Japan}
 \author{K. Nemoto}
 \affiliation{National Institute of Informatics, 2-1-2 Hitotsubashi, Chiyoda-ku, Tokyo 101-8430, Japan}
 \affiliation{Japanese-French Laboratory for Informatics, CNRS UMI 3527, 2-1-2 Hitotsubashi, Chiyoda-ku, Tokyo 101-8430, Japan}

\date{\today}

\begin{abstract}
\centering\begin{minipage}{\linewidth}

\noindent\textbf{Crystals arise as a result of the breaking of a spatial translation symmetry. Similarly, periodically-driven systems host non-equilibrium states when discrete symmetries in time are broken and time crystals arise. Here we introduce a novel method to describe, characterize and explore the physical phenomena related to this new phase of matter using tools from graph theory. 
This not only allows us to visualize time-crystalline order, but also to analyze features of the quantum system through the analysis of graphs. As an example, we explore in detail the melting process of a minimal model of a $2T$ Discrete Time Crystal ($2T$-DTC) and describe it in terms of the evolution of the associated graph structure. We show that during the melting process, the evolution of the network exhibits an emergent preferential attachment mechanism, directly associated with the existence of scale-free networks. Thus our strategy allows us to propose a new far-reaching application of time crystals as a quantum simulator of complex quantum networks.
}

\end{minipage}
\newline
\end{abstract}

\maketitle


\noindent {\textbf{INTRODUCTION}}

\noindent Symmetries are of utmost importance in condensed matter physics and statistical mechanics due to the strong relation between quantum phases of matter and symmetry breaking \cite{Shahar1997,Higgs1964,BCS1957,Luke1998}. Among the zoo of symmetry broken phases, discrete time crystals (DTC) play a fundamental role on their own due to the type of symmetry involved \cite{Sacha2018}. The time-crystalline phase, analogous to ``space'' crystals, arises when time (instead of space) translation symmetry is broken. The existence of quantum time crystals was originally proposed by Wilczek \cite{Wilczek2012}, and a discrete version of time crystals can be realized periodically-driven quantum systems \cite{Sacha2015,Else2016,Khemani2016}. It is in such cases that the system dynamics exhibits a subharmonic response with respect to the characteristic period of the drive caused by the synchronization in time of the particles of a many-body system \cite{Sacha2018}. 

The exploration of time crystals is a very active field of research and several experimental realizations with trapped ions \cite{Zhang2017}, dipolar spin impurities in diamond \cite{Choi2017}, ordered dipolar many-body systems \cite{Barrett2018}, ultracold atoms \cite{Smits2018} and nuclear spin-1/2 moments in molecules \cite{Pal2018} have been achieved. Yet an intuitive and complete insight of the nature of time-crystals and its characterization, as well as a set of proposed applications, is lacking. We here provide new tools based on graph theory and statistical mechanics to fill this gap. We propose a new strategy for the study and understanding of time symmetry-broken phases and their related phenomena \cite{Yao2017,Sacha2018-2}. As an example, we characterize the time-crystalline order and its melting. 

Our approach allows us to identify an instance of a perturbed time-crystal as a novel physical platform where to simulate complex networks whose evolution is governed by preferential attachment mechanisms \cite{Barabasi1999,Barabasi2002}. This type of networks, far from being regular or random, contains non-trivial topological structures present in many biological, social and technological systems. Small-world and scale-free networks are two of the most popular examples, the later being commonly characterized through power-law degree distributions that can be explained from the presence of a preferential attachment mechanism. The simulation of such networks has wide applicability, ranging from the study and understanding of behaviors present in communication or internet networks \cite{Yuan2012}, the development of new algorithms in deep learning \cite{Pal2018a}, or the analysis of genetic and neural structures in biological systems \cite{Dorogovtsev2003}.
\\
\newline


\noindent {\textbf{RESULTS}}\\
\noindent \textbf{Floquet theory in a nutshell}\\
The fundamentals of this work rely on the exploration of the dynamics of driven many-body quantum systems described by time-periodic Hamiltonians $\hat{H}(t+T)=\hat{H}(t)$, with $T=2\pi/\omega$ being the period of that drive. Our approach builds from the calculation of the Floquet operator $\hat{\mathcal{F}}=\hat{U}(T)=\mathcal{T}\exp(-\mathrm{i}\int_{0}^{T} \hat{H}(\tau) d\tau/\hbar)$,
which is the evolution operator within one period from time $t=0$ to time $t=T$~\cite{Haenggi1998,Polkovnikov2015,Anisimovas2015} with $\mathcal{T}$ being the time ordering operator.
In Floquet theory, one wishes to solve the eigenvalue problem $\hat{\mathcal{F}}\ket{\Phi_{s}}=e^{-\mathrm{i}\lambda_{s}T/\hbar}\ket{\Phi_{s}}$. The eigenvectors $\ket{\Phi_{s}}$  are known as Floquet states and $-\hbar\omega/2\leq\lambda_{s}\leq\hbar\omega/2$ are the quasienergies. At discrete times $t_n=nT$, the evolution can be understood in terms of an  effective ``time-independent" Hamiltonian $\hat{H}^{\text{eff}}$ such that $\hat{\mathcal{F}}=e^{-\mathrm{i}\hat{H}^{\text{eff}}T/\hbar}$. This Hamiltonian is very relevant in the context of Floquet engineering as it contains effective interactions that are absent in equilibrium systems, allowing it to be applied to quantum simulation problems \cite{Polkovnikov2015,Brandes2016,Eckardt2017}.
\newline
\\

\begin{figure*}[htbp]
  \centering
  \includegraphics[width=1\linewidth]{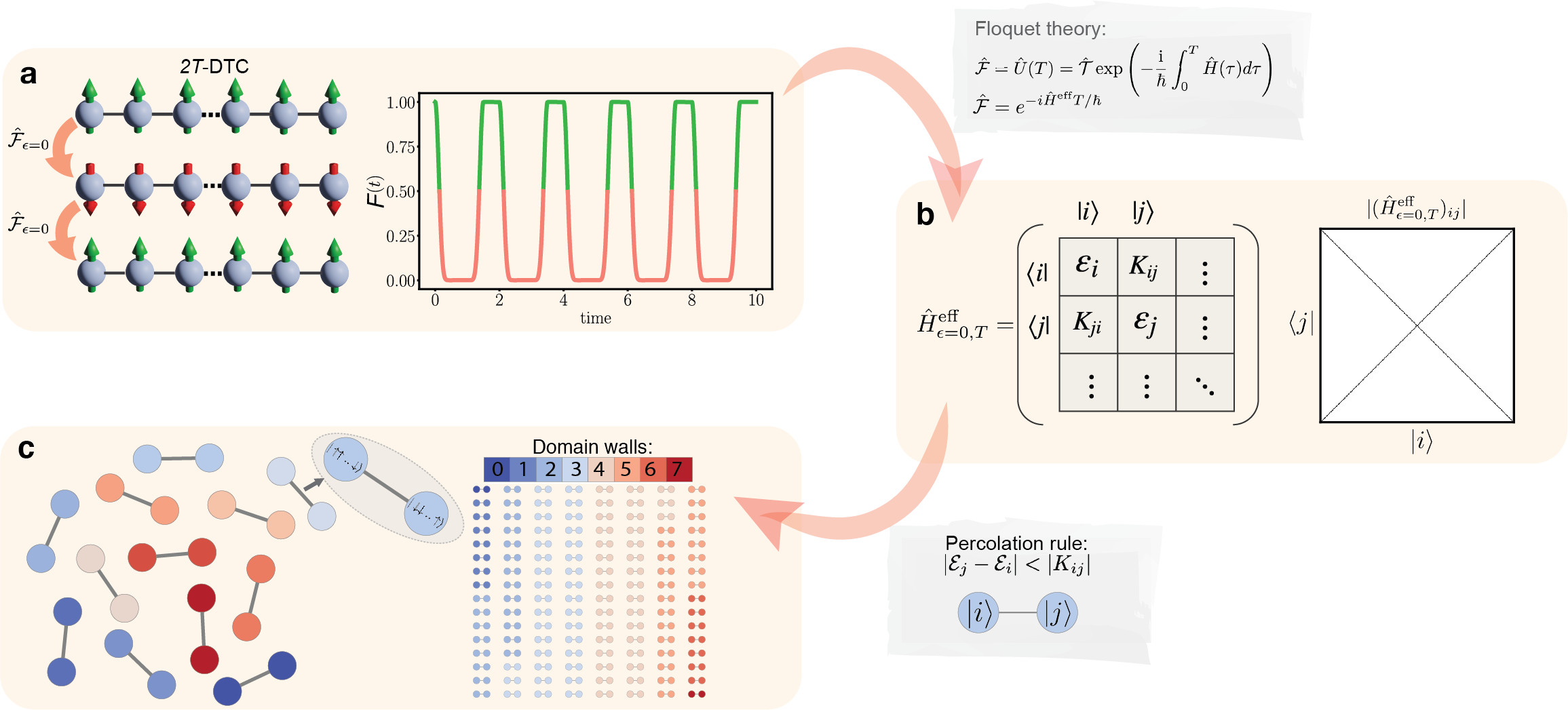}
  \caption{\textbf{Obtaining the associated graph of a $2T$-DTC.} \textbf{a}, On the left, diagram of the $2T$-DTC dynamics with no rotation error, $\epsilon$. The initial state ($\ket{\Psi(0)}=\ket{\uparrow\uparrow\uparrow...\uparrow}$), represented with the green arrows pointing up, is recovered after two periods of the driving protocol. From the first period, we obtain the unitary, $U(T)$, that will be used as the Floquet operator, $\hat{\mathcal{F}}_{\epsilon=0}$, to derive the effective Hamiltonian, $\hat{H}_{\epsilon=0,T}^{\text{eff}}$. On the right, fidelity of evolving state of an $n=8$ sites $2T$-DTC against its initial state, $F(t)=\vert\langle \Psi(0)\vert\Psi(t)\rangle\vert^2$ showing the $2T$ periodicity of the dynamics. \textbf{b}, The effective Hamiltonian, $\hat{H}_{\epsilon=0,T}^{\text{eff}}$, represented as a tight-binding matrix. $\mathcal{E}_i$ and $K_{ij}$ are the energies of the $\ket{i}$ configuration (see main text for full description) and transition energy between configurations $\ket{i}$ and $\ket{j}$, respectively. The right panel shows the entries of the effective Hamiltonian matrix, where only the diagonal and anti-diagonal entries are non-zero. \textbf{c}, After applying the percolation rule to the effective Hamiltonian we obtain an adjacency matrix, which is in turn represented as a graph with the nodes being each of the $2^n$ configuration basis set of the Hilbert space. In the right, for no rotation error, the crystal order of the $2T$-DTC can be observed as $2^{n-1}$ decoupled dimers. Here, the elements of each dimer are the configurations $\ket{i}$ and $\ket{2^n-1-i}$, which are related by a global $\pi$ rotation along the $x$-axis. All the nodes with same color have the same number of domain walls with the colormap gradient going from dark blue (0 domain walls) to dark red ($n-1$ domain walls).}
  \label{fig:diagram}
\end{figure*}

\begin{figure*}[ht!]
  \centering
  \includegraphics[width=1\linewidth]{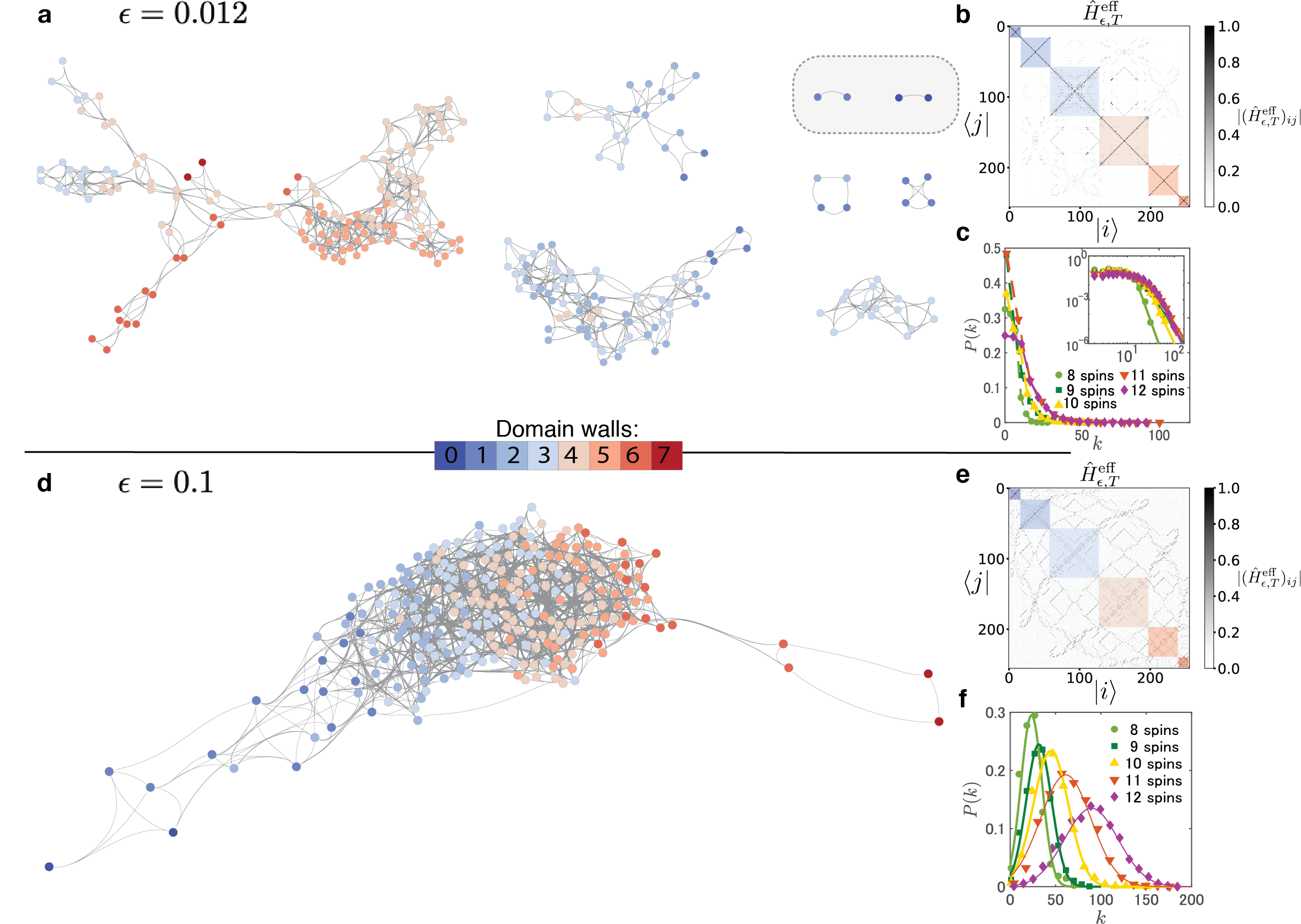}
 \caption{\textbf{Melting of a $2T$-DTC using $\hat{H}^{\text{eff}}_{\epsilon,T}$}. \textbf{a}, Graph representation obtained obtained from $\hat{H}^{\text{eff}}_{\epsilon,T}$ with $n=8$ sites and $\epsilon=0.012$. Nodes are color coded according to the number of domain walls of the corresponding configuration (see colormap). For moderate levels of error the nodes attach to each other according to their number of domain walls. This can also be observed in the effective Hamiltonian matrix of panel \textbf{b} with the basis ordered in increasing number of domain walls and delimited by the coloured squares. As the nodes start to cluster due to the presence of error, some non-zero off-diagonal terms appear in the centre of the matrix. For this level of error, some dimers survive (see top-right corner of panel \textbf{a}), serving as good indication of the robustness of the system and meaning that the crystal order is still present. In panel \textbf{c}, the degree distributions of the corresponding graph of panel \textbf{a} with different system sizes ($n=8-12$) averaged over 100 realizations of disorder are shown in both linear and logarithmic scale (see inset), which display heavy-tailed distributions. The distributions are fitted with a power-law curve (solid-lines in the inset), indicating the presence of large degree hub nodes. \textbf{d}, Graph representation obtained from $\hat{H}^{\text{eff}}_{\epsilon,T}$ with $n=8$ sites and $\epsilon=0.1$, and \textbf{e}. its associated effective Hamiltonian matrix. As we increase the error, the system forms a single cluster. This can be seen from the appearance of many new off-diagonal entries in the Hamiltonian matrix as well as the the presence of a giant component in the graph. In this scenario, the time crystal has melted completely and no crystal order is left (no presence of isolated dimers). The degree distributions, shown in \textbf{f}, also indicate that the heavy-tailed nature is destroyed, and approximate to a normal distribution.}
  \label{fig:TDTCgraph}
\end{figure*}

\begin{figure*}[ht!]
  \centering
  \includegraphics[width=1\linewidth]{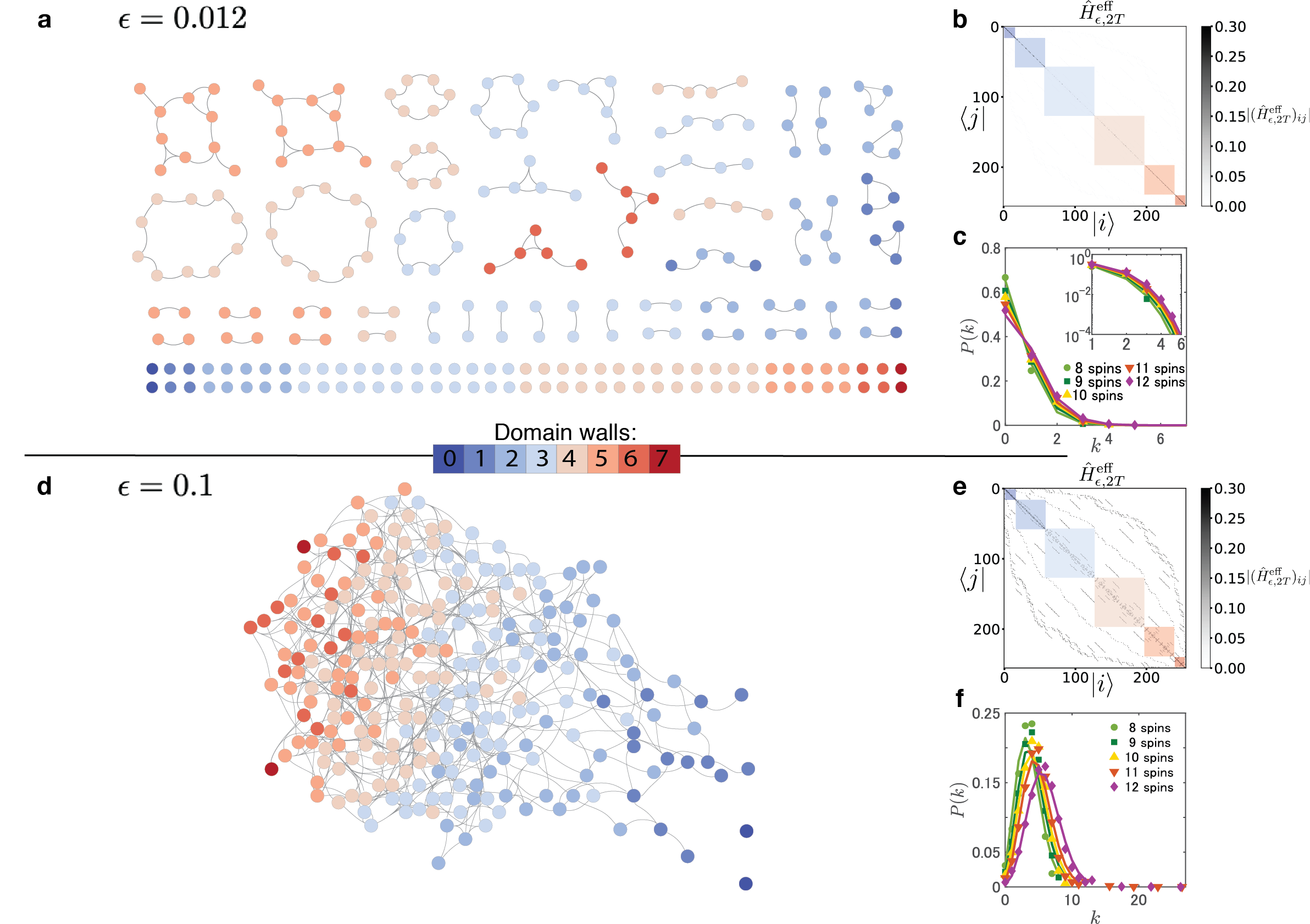}
  \caption{\textbf{Melting of a $2T$-DTC using $\hat{H}^{\text{eff}}_{\epsilon,2T}$}. \textbf{a}, Graph representation obtained from $\hat{H}^{\text{eff}}_{\epsilon,2T}$ with $n=8$ sites and a rotation error of $\epsilon=0.012$ with the nodes corresponding to configurations basis set of the Hilbert space color coded according to the number of domain walls of the corresponding state (see colormap). For none or moderate levels of error, the nodes are decoupled, as shown in the bottom part of the graph, or forming clusters of small size. In \textbf{b}, we present the effective Hamiltonian $\hat{H}^{\text{eff}}_{\epsilon,2T}$ that presents mainly diagonal terms. For moderate levels of error the nodes form small clusters. Again, they attach to each other according to their number of domain walls. In panel \textbf{c}, the degree distributions of the corresponding graph of panel \textbf{a} with different system sizes ($n=8-12$) averaged over 100 realizations of disorder are shown in both linear and logarithmic scale (see inset), fitted with a Poisson distribution. This indicates that the corresponding graph is a random graph with low connectivity. \textbf{d}, Graph representation obtained from $\hat{H}^{\text{eff}}_{\epsilon,2T}$ with $n=8$ and a rotation error of $\epsilon=0.1$. As we increase the error, the system forms a larger, highly-connected cluster. \textbf{e}, This can be seen from the increasing magnitude of the off-diagonal entries in the effective Hamiltonian matrix as well as the presence of a giant component in the graph. In this scenario, the time crystal has melted completely. \textbf{f}, The degree distributions are well approximated to a Poisson distribution in the lower degree region, which is the characteristic of highly connected random networks.}
  \label{fig:2TDTCgraph}
\end{figure*}


\newpage
\noindent \textbf{Discrete time crystals of period 2T}\\
We now move to a well-known Period-$2$ Discrete Time Crystal (2T-DTC) model \cite{Yao2017,Zhang2017} consisting of a one dimensional chain with $n$ spin-$1/2$ particles and governed by a time-dependent Hamiltonian with period $T = T_1 + T_2$,

 \begin{equation}
         \label{eq:DiscreteTimeCrystal}
  \hat{H}(t)=
  \begin{cases}
   \hat{H}_1 \equiv \hbar g \left(1 - \epsilon \right)\sum_{l} \sigma_{l}^{x} & 0 < t < T_1 \\
   \hat{H}_2 \equiv \hbar\sum_{lm}J_{lm}^{z} \sigma_{l}^{z} \sigma_{m}^{z} + \hbar \sum_{l}B_{l}^{z} \sigma_{l}^{z} & T_1 < t < T \ .
  \end{cases}
   \end{equation}
Here $\{\sigma_{l}^{x},\sigma_{l}^{y},\sigma_{l}^{z}\}$ are the usual Pauli operators on the $l$-th spin, $J_{lm}^{z} \equiv J_0/|l-m|^{\alpha}$ is the long-range interaction between spins $l,m$ that takes the form of an approximate power-law decay with a constant exponent $\alpha$, and $B_{l}^{z} \in \left[ 0,W \right]$ is a random longitudinal field. From the interplay between the driving, interactions and disorder, an MBL phase arises, being a key ingredient to allow the existence of DTC. Such a phase prevents the system from heating up to infinite temperature due to the effect of the drive and thus destroying the time-crystalline phase (see Sec. II of the Supplementary Material for a more detailed description). The parameter $g$ satisfies the condition $2gT_1=\pi$ such that $\hat{U}_1=\exp{\left( -\mathrm{i}\hat{H}_1 T_1/\hbar\right)}$ becomes a global $\pi$ pulse around the $x$-axis, with a rotation error $\epsilon$ also introduced. The Floquet operator is then given by:

\begin{equation}
         \label{eq:FloquetOperatorTimeCrystal}
         \hat{\mathcal{F}}_{\epsilon}=\hat{U}_{\epsilon}(T)=\exp{\left( -\frac{\mathrm{i}}{\hbar}\hat{H}_2 T_2\right)}\exp{\left( -\frac{\mathrm{i}}{\hbar}\hat{H}_1 T_1\right)}
         \ .
\end{equation}
Crucially, the Floquet operator depends on the error $\epsilon$, which determines how the time crystal will melt. In our work, we choose $\alpha=1.51$, $J_0T_2=0.06$ with a disorder strength $WT_2=\pi$, which are similar values used in the recent experiment~\cite{Zhang2017}. The key feature of this unitary evolution is that at each period $T$, as shown in Fig.~\ref{fig:diagram}.\textbf{a}, the system evolves between the states $\ket{\uparrow\uparrow\cdots\uparrow}$ and $\ket{\downarrow\downarrow\cdots\downarrow }$, when the state is initialized at one of these two. This periodic evolution is robust against moderate errors $\epsilon$ in $\hat{H_1}$. 
\newline
\\


\noindent \textbf{Floquet graphs}\\
Let us now introduce the concept of a Floquet graph as the graphical representation of the effective Hamiltonian, $\hat{H}^{\text{eff}}_{\epsilon,T}$ \cite{Bastidas2018}, obtained as $\hat{H}^{\text{eff}}_{\epsilon,T}=i \hbar/T \log[\hat{\mathcal{F}}_{\epsilon}]$. By shifting to the graph theory framework, we can provide a visual and intuitive understanding of the system and use its well-established tools to understand physical phenomena. We discuss the application of percolation to unveil the complex dynamics of time-crystals, even though such ideas can be widely applicable to any periodically-driven (Floquet) system.

The effective Hamiltonian, $ \hat{H}_{\epsilon,T}^{\text{eff}}$, is defined on a finite and discrete Hilbert space (dimension $N_\mathcal{H}$), and can be written as a tight-binding model using the configuration basis states $\lbrace |i\rangle \rbrace$, as

 \begin{equation}
 \label{eq:TightBinding}
        \hat{H}^{\text{eff}}_{\epsilon,T}=\sum_{i} \mathcal{E}_{i}|i\rangle \langle i| + \sum_{i,j} K_{ij}|i\rangle \langle j|=\sum_{s} \lambda_{s}|\Phi_s\rangle \langle \Phi_s|,
 \end{equation}
where $\mathcal{E}_{i}$ is the energy of configuration $|i\rangle$, and $K_{ij}$ is the transition energy between configurations $|i\rangle$ and $|j\rangle$ (see Fig.~\ref{fig:diagram}.\textbf{b}). The Floquet states $|\Phi_s\rangle$ are linear combinations of the configurations $|i\rangle$. In general, the quasienergies $\lambda_{s}$ are complicated functions of the parameters $\mathcal{E}_{i}$ and $K_{ij}$. In this work, we consider a system with $n$ spin-$1/2$ particles and the configurations $\lbrace |i\rangle \rbrace$ with $i=0,1,\ldots,2^n-1$ can be chosen as the $N_\mathcal{H}=2^n$ elements of the computational basis, i.e. the product states of the eigenstates of $\sigma_l^z$. In general, any configuration can be written as $\ket{i}=(\sigma^{+}_1)^{j_1}\cdots(\sigma^{+}_n)^{j_n}\ket{\downarrow\downarrow\cdots\downarrow }$, where $i=(j_1j_2\ldots j_n)_2$ is the binary decomposition of the integer $i$ and $\sigma^{+}_l\ket{\downarrow}_l=\ket{\uparrow}_l$. By using this notation, the states are denoted as $\ket{0}=\ket{\downarrow\downarrow\cdots\downarrow }$ and $\ket{2^n-1}=\ket{\uparrow\uparrow\cdots\uparrow}$. The basis configurations have been chosen to satisfy the relation $\ket{2^n - 1 - i}=\sigma^x_1\sigma^x_2\cdots \sigma^x_n\ket{i}$, such that the states $\ket{i}$ and $\ket{2^n - 1 - i}$ are always linked by the global $\pi$ rotation from $\hat{H}_1$ in Eq.~\ref{eq:DiscreteTimeCrystal},
and will constitute the $2^n$ nodes of the graph. In addition, such nodes will be linked according to a percolation rule \cite{Roy2019}, where a non-weighted edge is considered to be active between nodes $i$ and $j$ only if the condition

 \begin{equation}
 \label{eq:PercolationRule}
        |\mathcal{E}_{j} - \mathcal{E}_{i}| < |K_{ij}|
 \end{equation}
is satisfied. This rule eliminates the off-resonant transitions, allowing to effectively visualize the significant transitions in the Hamiltonian (see Sec.\ref{methods:A} of the Methods for further detail). 

In order to clarify the percolation on the graph, we define clusters as subsets of nodes which are connected with a path. The graph is percolated when all nodes belong to a single cluster~\cite{Potter2015}. We want to stress that the nodes represent many-body states $\ket{i}$ and not physical spin sites $l$, so the graph spans through the full Hilbert --configuration-- space (see Fig.~\ref{fig:diagram}.\textbf{c}). This percolation rule was proven to be useful to detect many-body localization to thermal phase transitions in undriven spin systems~\cite{Roy2019}.
\newline
\\


\noindent \textbf{Time crystal graphs}\\
Under the framework of Floquet graphs, the dynamics of the described $2T$-DTC can now be investigated. The effective Hamiltonian $\hat{H}^{\text{eff}}_{\epsilon,T}$ derived from Eq.~\eqref{eq:FloquetOperatorTimeCrystal} can be represented as a graph using the percolation rule from Eq.~\eqref{eq:PercolationRule}. The structure of the graph when $\epsilon=0$ presents decoupled dimers (two connected nodes), a signature of $2T$-DTCs crystals, as shown in the example presented in Fig.~\ref{fig:diagram}.\textbf{c}. We note that this visual characterization of the crystal time order is not unique of $2T$-DTC and can be extrapolated to crystals of different periodicity. For example, $3T$-DTC presents decoupled trimers; $4T$-DTC presents decoupled tetragons, and so on.  Our results hence reveal that, in the case of $2T$-DTCs and in the absence of error, there are decoupled two-dimensional subspaces. For example if there is an active link between the configurations $\ket{0}$ and $\ket{2^n-1}$ defined above, there is an energy gap $\hbar\pi/T$ between the quasienergies $\lambda_0$ and $\lambda_{2^n-1}$. The corresponding Floquet states are given by $\ket{\Phi_{0}}\approx(\ket{0}+\ket{2^n-1})/\sqrt{2}$ and $\ket{\Phi_{2^n-1}}\approx(\ket{0}-\ket{2^n-1})/\sqrt{2}$, which are maximally entangled GHZ states. The same behaviour is observed for other configurations, $\vert i\rangle$ and $\vert 2^n-1-i\rangle$, as all the paired Floquet states $\ket{\Phi_i}$, $\ket{\Phi_{2^n-1-i}}$ have an energy difference of $\hbar\pi/T$ \cite{Khemani2016}.
\newline
\\


\noindent \textbf{Melting of the $2T$-DTC}\\
Consider now the case where our system is initialised to $\ket{\psi(0)}=\ket{2^n-1}=\ket{\uparrow\uparrow\cdots\uparrow}$~\cite{Zhang2017}. This state is a superposition of Floquet states, $\ket{\psi(0)}\approx(\ket{\Phi_{0}}-\ket{\Phi_{2^n-1}})/\sqrt{2}$ and in the graph is represented as a dimer between the nodes $\ket{0}$ and $\ket{2^n-1}$. Such a dimer remains decoupled and isolated for modest levels of error (up to a ${\sim}3\%$ error in the rotation added as a non-zero value of $\epsilon$ in $\hat{H}_{1}$, see supplementary video). In Fig.~\ref{fig:TDTCgraph} we present the effect of moderate ($\epsilon=0.012$) and large ($\epsilon=0.1$) levels of rotation errors $\epsilon$ for a $2T$-DTC graph with $n=8$ spins. When the error increases there are couplings between the different subspaces and larger clusters appear in the graph (Fig.~\ref{fig:TDTCgraph}.\textbf{a}). If we keep increasing the error, the crystal will melt and its associated graph will take the form of a single highly --percolated-- connected cluster (Fig.~\ref{fig:TDTCgraph}.\textbf{d}). The presence/no presence of such a dimer indicates whether the system is still in a crystalline phase or, otherwise, has been melted. This demonstrates how our graph strategy, among other things, can track the robustness of the system in a very visual and intuitive manner. Importantly, we note that not all dimers show the same robustness, and some of them survive to higher error values. This, in fact, allows us to identify the most convenient state initialization: the initial state corresponding to one of the configurations of the most robust dimer will make the crystal phase last longer.
\newline
\\


\noindent \textbf{Conserved quantities}\\
Let us now analyze what factors are determinant in the robustness of each pair of configurations or, alternatively, how they cluster as the crystal melts. With our system represented as a graph, we can now focus on how to interpret its structural properties in terms of conserved quantities. To do so, it is of fundamental importance to note that the system is switching stroboscopically between two configurations, $\ket{i}$ and $\ket{2^n-1-i}$. If we prepare the system in a symmetry broken state in terms of its total magnetization and observe the system every two periods ($2T$), the system will effectively remain in a manifold of states with a fixed quasienergy. For no error, the graph will be formed by decoupled single nodes. In that sense, if we sample the dynamics every two periods, the system is in an effective equilibrium state determined by conserved quantities that are defined stroboscopically. Now let us construct the conserved quantities in the absence of error and show how they are destroyed when clusters in the graph are formed (Fig.~\ref{fig:2TDTCgraph}). 

To calculate the conserved quantities, let us consider the square of the Floquet operator of Eq.~\eqref{eq:FloquetOperatorTimeCrystal}, where $\hat{H}^{\text{eff}}_{\epsilon,2T}$ is the effective Hamiltonian now over two periods ($\hat{H}^{\text{eff}}_{\epsilon,2T}=i\hbar/2T\log[\hat{\mathcal{F}}^2_{\epsilon}]$). In that stroboscopic framework ($2T$), the Floquet operator for no error can be reduced to the following expression: 

\begin{equation}
         \label{eq:FloquetOperatorSquareNoError}
         \hat{\mathcal{F}}^2_{\epsilon=0}=\exp{\left( -2iT_2\sum_{lm}J_{lm}^{z} \sigma_{l}^{z} \sigma_{m}^{z} \right)}
         \ .
\end{equation}
In the absence of error, $\epsilon=0$, the disorder cancels out exactly and the effective Hamiltonian reads $\hat{H}_{\epsilon=0,2T}^{\text{eff}}=\hbar/2\sum_{lm}J_{lm}^{z} \sigma_{l}^{z} \sigma_{m}^{z}$. Despite its apparent simplicity, the effective Hamiltonian contains very relevant information. In particular, it preserves the local magnetization from which it follows the preservation of the total number of domain walls of the basis states $\hat{\mathcal{N}}=\sum_l(1-\sigma_{l}^{z} \sigma_{l+1}^{z})$ and parity $\hat{\Pi}=\prod_l \sigma^x_l$ such that $[\hat{H}_{\epsilon=0,2T}^{\text{eff}},\hat{\mathcal{N}}]=[\hat{H}_{\epsilon=0,2T}^{\text{eff}},\hat{\Pi}]=0$. Therefore our conserved quantities are classified according to parity and the total number of domain walls. If we prepare the system in the initial state $\ket{\psi(0)}=\ket{2^n-1}$, the system will remain stroboscopically within the symmetry multiplet with period-2 quasienergy $\lambda^{2T}_{1}=\hbar/2\sum_{lm}J_{lm}^{z}$ and zero domains walls $\langle\hat{\mathcal{N}}\rangle=0$. As shown in Fig.~\ref{fig:diagram}.\textbf{c}, in the absence of error, all the dimers conserve the number of domain walls (represented as a colormap) and their quasienergies. The aforementioned initial state, however, breaks the spatial parity symmetry $\hat{\Pi}$, which leads to a non-zero value of the correlation function $\langle\sigma^z_l(\tau)\rangle=\langle \psi(0)|\sigma^z_l(\tau)\sigma^z_l(0)|\psi(0)\rangle
\neq 0$, as observed in the experiment~\cite{Zhang2017}.
\newline
\\


\noindent \textbf{Breaking symmetries}\\
For zero error, the $2^n$ quasienergies are local integrals of motion and the Hilbert space can be classified by using the symmetries that preserve the aforementioned conserved quantities. In turn, the effective Hamiltonian $\hat{H}^{\text{eff}}_{\epsilon,2T}$ is block diagonal (see Fig.~\ref{fig:2TDTCgraph}.\textbf{b}) with $n$ sub-blocks given by the conservation of the number of domain walls. In the absence of error $(\epsilon=0)$, and after $2T$, one can find a classical quasienergy surface in phase space associated to the effective Hamiltonian $\hat{H}_{\epsilon=0,2T}^{\text{eff}}$. The idea is to represent the spins as vectors in a three-dimensional space. By using polar coordinates, one can derive the classical Hamiltonian in a $n$-dimensional phase space
\begin{equation}
      \label{eq:ClassicalEnergy}
            \mathcal{H}_{\epsilon=0,2T}^{\text{eff}}(\theta_{1},\theta_{2},\ldots,\theta_{n})=\frac{\hbar}{2}\sum_{lm}J_{lm}^{z} \cos\theta_{l} \cos\theta_{m}
            \ ,
\end{equation}
where $\theta_l$ is the polar angle. From this one can show that for zero error $\epsilon=0$ all configurations are stable fixed points. The presence of the error brings the system far from the classical limit and quantum effects become dominant. In such scenario the configurations are not fixed points of the dynamics anymore and they become unstable (see Sec. III and IV of the Supplementary Material for more detail). To investigate this, we use the Baker-Campbell-Hausdorff formula to obtain, up to a first order approximation, the effective Hamiltonian for a non-zero value of $\epsilon$:

\begin{align}
      \label{eq:EffectiveHamiltonianNonZeroError}
            \hat{H}_{\epsilon,2T}^{\text{eff}}&=\frac{\hbar T_2}{T}\sum_{lm}J_{lm}^{z}\sigma_{l}^{z} \sigma_{m}^{z}-\frac{\hbar g\epsilon T_1}{2T}\sum_{l}[(\cos{(B_l2T_2)}+1) \sigma_{l}^{x}
            \nonumber\\&
            +\sin{(B_l2T_2)} \sigma_{l}^{y}] \ .
\end{align}
This allows us to understand how $\epsilon$ destroys the symmetries by coupling different symmetry multiplets. This can be observed from Fig.~\ref{fig:2TDTCgraph}.\textbf{b}-\textbf{e}, as off-diagonal entries start populating the effective Hamiltonian matrix due to the error. This situation resembles the Kolmogorov-Arnold-Moser (KAM) theory in classical mechanics, which describes how invariant tori are broken in phase space under the effect of perturbations \cite{Lazutkin1993}, with our perturbation being $\epsilon$. First of all, the transverse field term $-\hbar g \epsilon/4\sum_{l}\sin{(B_lT)} \sigma_{l}^{y}$ breaks the parity symmetry. Most importantly, the transverse field term $-\hbar g \epsilon/4\sum_{l}(\cos{(B_lT)}+1)\sigma_{l}^{x}$, while preserving parity, melts the ferromagnetic order and breaks the conservation of the number of domain walls. One of the most interesting aspects of our work is the striking similarity with KAM theory in classical systems: a small perturbation can destroy integrals of motion \cite{Lazutkin1993}. The most fragile integrals of motion are periodic unstable orbits in phase space as discussed in the context of Eq.~\eqref{eq:ClassicalEnergy}. This explains the robustness of the time crystal and how clusters form in the graph, as we increase the error. The time crystal is robust against small errors because the last integral of motion that is destroyed by the perturbation is the configuration with the lowest number of domain walls: this is, $\ket{\Psi(0)}=\ket{2^n-1}$ (see darkest blue dimer on the top-right corner of Fig.~\ref{fig:TDTCgraph}.\textbf{a}). From now on we will use this model as an example to illustrate the applicability of our method.
\newline
\\


\noindent \textbf{Preferential attachment}\\
Let us now put the focus on the network topology for moderate levels of rotation error ($\epsilon\sim0.012$). In the previous section we have seen that the quantum terms that appear in Eq.~(\ref{eq:EffectiveHamiltonianNonZeroError})
break the conservation of number of domain walls. The error lifts the degeneracies present in the spectrum and new transitions between close states appear. In terms of the graph, the nodes with the same or similar number of domain walls connect following a preferential attachment mechanism. As most nodes in the network have $n/2$ or $n/2 - 1$ ($n$: even) number of domain walls, and they have very close quasienergies, these nodes easily connect to each other with a small value of error. This leads to the appearance of the large degree hub nodes as well as the heavy tailed degree distributions shown in Fig.~\ref{fig:TDTCgraph}.\textbf{c}. The tail of these distributions can be fit to a power law distribution, which is the characteristic of scale-free networks. Following Clauset \textit{et al.}'s recommendations~\cite{clauset2009power}, we further test the goodness of fit of the power law against the lognormal distribution for each of these distributions, and observe that the power law favours over the lognormal distribution (see Sec. \ref{methods:B} of the Methods for further detail).

\begin{figure}
    \centering
    \includegraphics[width=\linewidth]{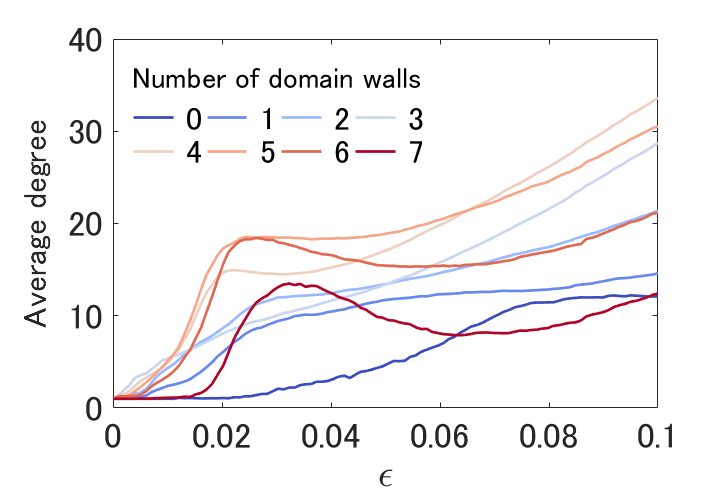}
    \caption{The average degree of the graph obtained from $\hat{H}^{\text{eff}}_{\epsilon,T}$ with $n=8$ sites, plotted against the error $\epsilon$. The nodes are distinguished by their number of domain walls, and averaged individually for each values of domain walls, as well as averaging over $100$ realizations of disorder. Notice that some of the curves exhibit a local maximum approximately between $0.01<\epsilon<0.03$. This is exactly the range in which we observe the scale-free behavior emerging in our system. Comparing each curves also confirms that the states with lower number of domain walls are more robust against the error in a sense that they have lower degrees.}
    \label{fig:degreefepsilon}
\end{figure}
In addition, exploring the behaviour of the average degree of the graph obtained from $\hat{H}^{\text{eff}}_{\epsilon,T}$ further captures the existence of the preferential attachment mechanism and the non-trivial melting process of the $2T$-DTC. From Fig.~\ref{fig:degreefepsilon}, we can confirm that the nodes with 5 or 6 domain walls (where the $n=8$ case is considered here) tend to acquire neighbours preferentially and have more degree already from a small error $\epsilon < 0.02$. This is the region where we can observe the scale-free behaviour emerging. When the error is increased, the preferential behaviour becomes weaker, and eventually most of the nodes tend to have large degree, closer to the structure of random networks, which indicates the crystal has melted. 

This behaviour not only has consequences for the dynamics of the time crystals but can also be used to perform quantum simulation of networks with exotic properties such as scale-free-like ones \cite{Barabasi2002}. Note that the scale-free-like networks shown in Fig.~\ref{fig:TDTCgraph} are only observed in the effective Hamiltonian obtained from single period $\hat{H}^{\text{eff}}_{\epsilon,T}$ and not in the one from two periods $\hat{H}^{\text{eff}}_{\epsilon,2T}$, which shows Poisson distributions, typical for random networks \cite{Migdal2013,Barabasi2002}. We observe that a small error has little effect to the effective Hamiltonian $\hat{H}^{\text{eff}}_{\epsilon,2T}$ (as shown in Fig.~\ref{fig:2TDTCgraph}.\textbf{b)}). This can be understood from the fact that each $\pi$-rotation changes the direction of the rotation around the z-axis caused by the $B^z$ term. Therefore at time $2T$ the effect of the $B^z$ term is effectively cancelled, limiting the number of transitions happening. On the other hand, this is not the case at time $T$, and off-diagonal terms appear in the matrix form of $\hat{H}^{\text{eff}}_{\epsilon,T}$ as shown in Fig.~\ref{fig:TDTCgraph}.\textbf{b)}. This indicates that transitions to a larger fraction of states can occur, which in turn yields a scale-free-like network for moderate values of $\epsilon$.
\newline
\\


\noindent {\textbf{CONCLUSION}}\\
Our approach of representing discrete time crystals in terms of graphs is key to understand the structure of such an exotic phase of matter and to envisage prospective applications. We explicitly show that by translating a $2T$-DTC into a graph theory language we can study in detail how the crystal order disappears as an increasing error melts the time crystal. Among others, it allowed us to identify the crucial role symmetries and conserved quantities play in the resilience of the crystal, by observing the preferential attachment mechanism present in the formation of clusters for this specific model of time crystal. Crucially, the nature of the obtained graphs for moderate levels of error suggests that such systems could be used as scale-free-like network simulators, giving rise to a novel application of such devices in the field of complex networks \cite{Paparo2013,Faccin2014}. Importantly, because our networks span the configuration space, such simulation could be done with moderate numbers of qubits and thus available noisy intermediate-scale quantum (NISQ) platforms (ranging from ion traps to superconducting qubit chips) could be used for its implementation. Structural information about the network (i.e. the degree distribution) using DTCs in NISQ devices could be experimentally obtained by exploiting a quantum walk in the configuration space. One could track down the number of configurations visited dynamically in a given time by measuring a set of $l$-point spin correlation functions, a measure that is experimentally accessible with current technology. The number of configurations visited in a given time for different initial conditions is related to the degree of the different nodes of the graph, unveiling the complexity of the network (see Sec. V of the Supplementary Material for more detail on our proposed experimental protocol). Future work will include further investigation and characterization on the nature of these networks as well as the study of additional phenomena present in time crystals in terms of graphs. We believe that the use of this formalism will lead not only to a more complete and deep understanding of discrete time crystals and their related phenomena, but also be very advantageous in the study of periodically driven quantum systems.
\newline
\\


\textit{Acknowledgements:\textemdash}
We thank A. Sakurai, M. Hanks, T. Haug, Y. Naka Renoust, and J. Schmiedmayer for valuable discussions. This work was supported in part from the Japanese program Q-LEAP, the MEXT KAKENHI Grant-in-Aid for Scientific Research on Innovative Areas Science of hybrid quantum systems Grant No.15H05870 and the JSPS KAKENHI Grant No. 19H00662. This project was also made possible through the support of a grant from the John Templeton Foundation (JTF 60478). The opinions expressed in this publication are those of the authors and do not necessarily reflect the views of the John Templeton Foundation. Author contributions: M.P.E, T.O. and V.M.B conceived the initial idea of the research. All authors contributed to develop the idea and analyse the system.  M.P.E, T.O., V.M.B and B.R. performed numerical simulations. All authors contributed to discussions of the results and the development of the manuscript. K.S., W. J. M. and K.N. supervised the whole project. The authors declare that they have no competing interests. Data Availability: All data needed to evaluate the conclusions in the paper are present in the paper and/or the Supplementary Materials. Additional data related to this paper may be requested from the authors. \newline
\\


\bibliographystyle{unsrt}
\bibliography{MBL,Mybib}


\section*{Methods}

\subsection{Floquet theory: Percolation rule and graph structure}
\label{methods:A}
\noindent

One of the main tools used in our work is the percolation rule, which establishes when there is a link between two nodes of a graph representing the effective Hamiltonian. In this section we explain in detail the motivation of the percolation rule and its interpretation in terms of Floquet theory. In order to have an intuitive picture of the percolation rule, let us consider only two configurations $\ket{i}$ and $\ket{j}$ with energies $E_i$ and $E_j$ in the absence of drive, respectively. For simplicity, let us assume that the drive induces transitions only between the aforementioned configurations. Within the two-state approximation, the Hamiltonian from Eq.\eqref{eq:DiscreteTimeCrystal} reads:

\begin{equation}
         \label{eq:TwoLevSystem}
  \hat{H}(t)=
  \begin{cases}
   \hat{H}_1 \equiv \hbar g \left(1 - \epsilon \right)\left(|i\rangle \langle j|+|j\rangle\langle i|\right) & 0 < t < T_1 \\
   \hat{H}_2 \equiv  E_{i}|i\rangle \langle i| +  E_{j}|j\rangle \langle j| & T_1 < t < T_2 \ .
  \end{cases}
   \end{equation}

In general, the transitions occur when the driving is resonant with the energy level spacing between two configurations, i.e., when $E_j-E_i=\hbar k\omega$, where $k$ is an integer and $\omega=2\pi/T$ is the frequency of the drive. Therefore, when the drive is switched on, the quasienergies $\lambda_{s_1}$ and $\lambda_{s_2}$ corresponding to the energies $E_i$ and $E_j$, exhibit an anticrossing~\cite{Haenggi1998}.
In the language of Floquet theory, the latter is referred to as a $m$-photon resonance because the quasienergies are defined modulo $\omega$.

If we restrict ourselves to the configurations $\ket{i}$ and $\ket{j}$, the effective Hamiltonian can be written as
\begin{eqnarray}
 \label{eq:TightTwoLev}
        \hat{H}^{\text{eff}}_{\epsilon,T}=\mathcal{E}_{i}|i\rangle \langle i|+\mathcal{E}_{j}|j\rangle \langle j| +  K_{ij}|i\rangle \langle j|+K_{ji}|j\rangle \langle i|= \\ \lambda_{s_1}|\Phi_{s_1}\rangle \langle \Phi_{s_1}|+\lambda_{s_2}|\Phi_{s_2}\rangle \langle \Phi_{s_2}|
        \ .
 \end{eqnarray}

From this we can see that the energy gap for the quasienergies scales as $\delta\lambda=|\lambda_{s_1}-\lambda_{s_2}|=\sqrt{(\mathcal{E}_{i}-\mathcal{E}_{j})^2+(K_{ij})^2}$. In addition, we can derive the expressions for the eigenstates $\ket{\Phi_{s_1}}=\cos\theta_{ij}\ket{i}+\sin\theta_{ij}\ket{j}$ and $\ket{\Phi_{s_2}}=-\sin\theta_{ij}\ket{i}+\cos\theta_{ij}\ket{j}$, where $\cos\theta_{ij}=|\mathcal{E}_{i}-\mathcal{E}_{j}|/\delta\lambda$ and $\sin\theta_{ij}=|K_{ij}|/\delta\lambda$. The percolation rule $|\mathcal{E}_{j} - \mathcal{E}_{i}| < |K_{ij}|$ from Eq.~\eqref{eq:PercolationRule} implies that when the link between the $i$-th and $j$-th nodes is active, the Floquet states $\ket{\Phi_{s_1}}$ and $\ket{\Phi_{s_2}}$ with quasienergies $\lambda_{s_1}$ and $\lambda_{s_2}$ are delocalized in the basis $\ket{i}$ and $\ket{j}$. On the contrary, if the percolation rule is violated, i.e., $|\mathcal{E}_{j} - \mathcal{E}_{i}| > |K_{ij}|$, the Floquet states are localized and no link between the $i$-th and $j$-th nodes is drawn in the graph. 

Now let us consider a more detailed derivation of the percolation rule by using a locator-like expansion for driven systems. Here we explicitly derive the percolation rule and motivate its origin in terms of resonances. Let us first shortly discuss the most relevant aspects of the locator expansion. In the Anderson model, one can expand the resolvent in terms of the hopping strength. For strong disorder, this can be understood as a perturbative expansion of the localized wavefunctions in the hopping term of the Hamiltonian, which is referred to as the locator expansion. The latter usually diverges due to resonances~\cite{Anderson1958}.

In driven quantum systems, the Hamiltonian is periodic in time and all the important information is contained in the Floquet operator $\hat{\mathcal{F}}=\hat{U}(T)$, which is the evolution operator in a period $T$ of the drive. Let us begin be defining the discrete transformation
\begin{equation}
          \label{eq:Ztransform}
          \hat{G}_{\text{F}}(z)=\sum^{\infty}_{l=0}\hat{\mathcal{F}}^n e^{\mathrm{i}nz}=\frac{1}{1-\hat{\mathcal{F}}e^{\mathrm{i}z}}
          \ .
 \end{equation}
 
The operator $\hat{G}_{\text{F}}(z)$ now plays the role of the resolvent of the Floquet operator $\hat{\mathcal{F}}$.  Next we can express the Floquet operator in terms of the effective Hamiltonian, as $\hat{\mathcal{F}}=e^{-\mathrm{i} \hat{H}^{\text{eff}}_{\epsilon,T} T} $, where $ \hat{H}^{\text{eff}}_{\epsilon,T}=\hat{H}_0+\hat{V}$ with $\hat{H}_0=\sum_i \mathcal{E}_{i}|i\rangle \langle i|$ and $\hat{V}=  \sum_{i,j}K_{ij}|i\rangle \langle j|$.

Although Eq.\eqref{eq:Ztransform} look different than the resolvent for undriven systems~\cite{Anderson1958}, they share the same structure if we consider the expansion
\begin{align}
          \label{eq:ZtransformExpansion}
          \hat{G}_{\text{F}}(z)&=\sum^{\infty}_{m=0}\frac{B_m (-\mathrm{i}T)^{m-1}}{m!}   (z-\hat{H}^{\text{eff}}_{\epsilon,T})^{m-1} 
          \nonumber\\
          &\approx  \frac{\mathrm{i}}{T(z-\hat{H}^{\text{eff}}_{\epsilon,T})} +B_1+\cdots   
 \end{align}
in terms of the generating function for the Bernoulli numbers $B_m$~\cite{abramowitz1988handbook}.
 
The next step is to use the perturbative expansion of Eq.~\eqref{eq:ZtransformExpansion} in the hopping term $\hat{V}$ 
\begin{align}
          \label{eq:FloquetResolventExpansion}
      \frac{1}{z-\hat{H}^{\text{eff}}_{\epsilon,T}}=\frac{1}{z-\hat{H}_0}+\left(\frac{1}{z-\hat{H}_0}\right)\hat{V}\left(\frac{1}{z-\hat{H}_0}\right)+\cdots
      \ .
 \end{align}
This allows us to calculate the transition probability between two configurations $\ket{i}$ and $\ket{j}$, as follows
\begin{equation}
          \label{Eq:TransitionProb}
          \bra{i}\hat{G}_{\text{F}}(z)\ket{j}\approx\frac{\mathrm{i}\delta_{i,j}}{T(z-\mathcal{E}_{i})}+\frac{\mathrm{i} K_{ij}}{T(z-\mathcal{E}_{i})(z-\mathcal{E}_{j})}+B_1\delta_{i,j}+\cdots
          \ ,
 \end{equation}
where $\delta_{i,j}$ is the Kronecker delta. 
 To investigate the physical meaning of the percolation rule, let us consider the role of the complex variable $z=\varepsilon+\mathrm{i}\delta$ when $\varepsilon=\mathcal{E}_{i}$. In this case, we can see that the relevant information of the series expansion is given by the ratio $|K_{ij}|/|\mathcal{E}_{i}-\mathcal{E}_{j}|$. This resembles the locator expansion for the undriven Anderson model. The main difference here is that that  $\ket{i}$ and $\ket{j}$ are not lattice sites but configurations of spins. In any case, the series expansion \eqref{Eq:TransitionProb} exhibit divergences whenever the condition $|\mathcal{E}_{i}-\mathcal{E}_{j}|<|K_{ij}|$ is satisfied. Keeping in mind  the explanation of the percolation rule in terms of two level systems, here the divergence is a signature of "hibridization" of the configurations $\ket{i}$ and $\ket{j}$. Thus, when the configurations are close to resonance, the states become delocalized under the action of the hopping term. On the contrary, when $|\mathcal{E}_{i}-\mathcal{E}_{j}|>|K_{ij}|$, the states remain localized.

\subsection{Power law fitting of the degree distributions}
\label{methods:B}

Let us first define the concept of degree within the context of graph theory. The degree of a vertex v belonging to a graph G is the number of edges connected to that vertex v. The degree distribution is therefore a probability distribution of vertex degrees over the whole graph G. Here we explain how the degree distributions of the graphs obtained from the effective Hamiltonian of the $2T$-DTC can be fit to a power law function $p(k) \propto k^{-\beta}$. We have followed the method by Clauset \textit{et al.}~\cite{clauset2009power}. First, the degree distributions $P(k)$ are obtained by measuring log-binned histograms of the degrees $k$ of the graph. We have obtained the distributions from 100 realizations of the graph with different disorder values for statistical convergence. Generally, the power law behaviour is observed in the large degree tail of the distribution rather than the entire domain. Therefore, we estimate the lower bound $k = k_{\text{min}}$ where the best power law fit can be obtained. This is done by estimating the power law exponent $\beta$ by applying maximum likelihood method on a certain portion $k \geq k_{\text{cutoff}}$ of the distribution. Once the exponent is estimated, the Kolmogorov-Smirnov distance between the distribution and the fit is computed. We compute this with various $k_{\text{cutoff}} > 0$ and the $k_{\text{min}}$ is chosen at the $k_{\text{cutoff}}$ where the Kolmogorov-Smirnov distance is minimized.
Once the $k_{\text{min}}$ and $\beta$ is estimated, we examine the goodness of fit of the power law using a comparative test. We consider an alternative probability distribution function that may be good fit for our data, and apply a likelihood ratio test between the power law and the alternative function. Here we chose the lognormal function as the alternative, which is another heavy-tailed function. From this comparative test, we conclude that the power law function well describes the degree distributions of our graphs.


\pagebreak
\clearpage
\onecolumngrid

\renewcommand{\Re}{\mathrm{Re}}
\renewcommand{\Im}{\mathrm{Im}}

\newcommand{\beginsupplement}{%
        \setcounter{table}{0}
        \renewcommand{\thetable}{S\arabic{table}}%
        \setcounter{figure}{0}
        \renewcommand{\thefigure}{S\arabic{figure}}%
        \setcounter{equation}{0}
        \renewcommand{\theequation}{S\arabic{equation}}
        \setcounter{section}{0}
        \setcounter{subsection}{0}
        \setcounter{tocdepth}{-10}
     }

\renewcommand{\thefootnote}{\fnsymbol{footnote}}

\beginsupplement
\begin{center}
  \textbf{\large Supplementary materials for ``Simulating complex quantum networks with time crystals"}\\[.2cm]
  M. P. Estarellas,$^{1,}$\footnote[1]{These authors contributed equally.} T. Osada,$^{2,1,\red{*}}$, V. M. Bastidas$^{3,\red{*}}$, B. Renoust$^{4,1,5}$, K. Sanaka$^{2}$, W. J. Munro$^{3,1}$ and K. Nemoto$^{1,5}$\\[.1cm]
  {\itshape ${}^1$National Institute of Informatics, 2-1-2 Hitotsubashi, Chiyoda-ku, Tokyo 101-8430, Japan\\
  ${}^2$Tokyo University of Science, 1-3 Kagurazaka, Shinjuku, Tokyo, 162-8601, Japan\\
  ${}^3$NTT Basic Research Laboratories \& Research Center for Theoretical Quantum Physics,\\  
  3-1 Morinosato-Wakamiya, Atsugi, Kanagawa, 243-0198, Japan\\
  ${}^4$Osaka University, Institute for Datability Science, 2-8 Yamadaoka, Suita, Osaka Prefecture 565-0871, Japan\\
  ${}^5$Japanese-French Laboratory for Informatics, CNRS UMI 3527,\\
  2-1-2 Hitotsubashi, Chiyoda-ku, Tokyo 101-8430, Japan\\}
\end{center}

\openup 1em
\tableofcontents
\openup -1em

\section{A roadmap through the supplementary materials}
\addcontentsline{toc}{subsubsection}{I.  A roadmap through the supplementary materials}
The purpose of this supplementary material is to provide the reader with the necessary tools to understand the main manuscript. In section~\ref{SI2} we focus on the relation between the robustness of time crystals and many-body localization (MBL). We explain in detail how to obtain the level statistics for the quasienergy gaps, which is a well-known diagnostic of MBL. In section~\ref{SI3} we show how to obtain signatures of the melting process by calculating the power spectrum of a time series of measured observables. Our approach is experimentally feasible and provides clear evidence that some configurations are more stable than others. In section~\ref{SI4} we provide a semiclassical description of the stability of the different spin configurations. Finally, a very important aspect of our results is that there is an emergent scale-free behavior in the graphs associated to the DTC. With this purpose in mind, in section~\ref{SI5} we provide an experimental protocol to simulate complex quantum networks in near-term quantum devices. Our approach is based on a quantum walk in the configuration space, which can be realized by measuring $l$-point spin correlation functions.

\section{Level statistics of the quasienergies and manybody localization}
\addcontentsline{toc}{subsubsection}{II. Level statistics of the quasienergies and manybody localization}
\label{SI2}
In classical mechanics, integrable systems are characterized by the existence of integrals of motion restricting the motion to orbits that are constrained by conservation rules. The classical harmonic oscillator, for example, exhibits a two-dimensional phase space with coordinates $q$ and $p$ representing position and momentum, respectively. 
In the absence of driving, the energy of the oscillator is conserved. Thereby, given an initial condition $(q_0,p_0)$, the classical orbits are constrained to the energy surface  $E(q,p)=p^{2}/2M+M\Omega q^2/2=p_0^{2}/2M+M\Omega q_0^2/2$, where $M$ is the mass and $\Omega$ is the oscillation frequency of the oscillator. For any value of the energy, the classical orbits are periodic with a period $T_{\text{orbit}}=2\pi/\Omega$.
In the quantum world, quantum signatures of the periodic orbits appear in the spacings $\delta_n=E_{n+1}-E_{n}=\hbar\Omega$ between the eigenvalues $E_n=\hbar\Omega(n-1/2)$ of the Hamiltonian $\hat{H}=\hbar\Omega(\hat{a}^{\dagger}\hat{a}-1/2)$, with $n$ labelling each of the system's eigenstates and $\hat{a}$ ($\hat{a}^{\dagger}$) being the anhiliation (creation) operator. Interestingly, if we calculate the probability distribution of level spacings $\delta_n$, this would be a Dirac delta distribution centered at $\Omega$. When we break conservation rules, however, the integrability of the system is compromised because the conserved quantities are destroyed and the level spacing distribution largely deviates from a simple Dirac-delta type distribution.  This statistical analysis is one of the cornerstones of recent developments in the theory of thermalization, which is closely related to random matrix theory and quantum signatures of chaos~\cite{Srednicki1994, dalessio16}. 

Recently, there has been an increasing interest in the study of level statistics as a diagnosis for thermalization and manybody localization. In the context of quantum manybody systems, the interactions distribute the energy between all the available states, because they usually break conserved quantities at the single-particle level, leading to thermalization and ergodic behavior. In systems where there is a well defined semiclassical limit, this behavior is associated with a fully chaotic phase space~\cite{Srednicki1994, dalessio16}. As we discussed above, in quantum mechanics, level statistics is a powerful tool to determine the nature of the dynamics. In terms of level statistics, this means that the statistical behavior of an ergodic system will largely deviate from the quantum harmonic oscillator case described above.  In particular, the destruction of conserved quantities induces strong correlations between the energy eigenvalues that are reflected in a strong level repulsion~\cite{dalessio16}.  In the case of disordered manybody systems, there is a competition between disorder and interactions. Disorder can suppress ergodicity giving rise to the manybody localized phase, which is characterized by an extensive number of conserved quantities~\cite{nandkishore15,Vosk2015,khemani17}. In the absence of drive, one is interested in the distribution of ratios $P(r)$ with $r_{s}=\min(\delta_{s},\delta_{s+1})/\max(\delta_{s},\delta_{s+1}) \leq 1.0$ with $\delta_{s}=E_{s+1}-E_{s}$, where $E_{s}$ are the energy eigenvalues sorted in order of increasing value~\cite{dalessio16}. In the MBL phase, the levels are uncorrelated because the manybody states are highly localized in space and the level statistics follows a Poissonian distribution
\begin{equation}
\label{eq:PoissonDist}
P_{\text{Poisson}}(r)=\frac{2}{(1+r)^2}
\ .
\end{equation}
When the interactions overcome the effect of disorder, the aforementioned conserved quantities are destroyed and the system is able to explore all the configuration space with a constant energy. This is reflected in a strong level repulsion because some states are delocalized. As a consequence, in the case of real Hamiltonians, the system follows a universal level statistics
\begin{equation}
\label{eq:GOEDist}
P_{\text{GOE}}(r)=\frac{27}{4}\frac{r+r^2}{(1+r+r^2)^{5/2}}
\end{equation}
associated to the Gaussian Orthogonal ensemble (GOE) of random matrices~\cite{dalessio16,roushan17}.

In our manuscript, we are interested in the dynamics of DTCs under the effect of a rotation error as shown in the Hamiltonian from Eq. (1). Below we show that the time-crystalline order is directly related to the presence of an MBL phase. In analogy to the quantum harmonic oscillator, in the absence of error, the time crystal has several conserved quantities and the motion is periodic. However, discrete time crystals are quantum phases of matter that appear in periodically-driven quantum systems and energy is not conserved anymore. Therefore,  we have to work with gaps $\delta_{s}=\lambda_{s+1}-\lambda_{s}$, where $-\hbar\omega/2<\lambda_{s}<\hbar\omega/2$ are the quasienergies and $\omega$ is the frequency of the drive. Under the effect of a small rotation error, there is a small coupling between different symmetry multiplets and the quasienergy level statistics follows a Poissonian behavior as in Eq.~\eqref{eq:PoissonDist}. As a consequence, the system is in the MBL phase, which protects the system from heating up to infinite temperatures~\cite{Rigol2014,Khemani2016}. When the error is increased, the conserved quantities are destroyed and the Floquet states become highly delocalized in the configuration space. The DTC melts and an MBL-to-ergodic phase transition takes place. The statistics of levels in this case is given by

\begin{equation}
\label{eq:COEDist}
P_{\text{COE}}(r)=\frac{2}{3}\left\{\left[\frac{\sin\left(\frac{2\pi r}{r+1}\right)}{2\pi r^2}\right]+\frac{1}{(1+r)^2}+\left[\frac{\sin\left(\frac{2\pi}{r+1}\right)}{2\pi }\right]\right\}
-\frac{2}{3}\left\{\left[\frac{\cos\left(\frac{2\pi }{r+1}\right)}{2\pi r^2}\right]+\left[\frac{\cos\left(\frac{2\pi r}{r+1}\right)}{r(r+1)}\right]\right\}
\,
\end{equation}
which is the same statistics as the circular orthogonal ensemble of random matrices~\cite{Rigol2014,tangpanitanon2019quantum}. 
\begin{figure}
	\centering
	\includegraphics[width=0.95\textwidth]{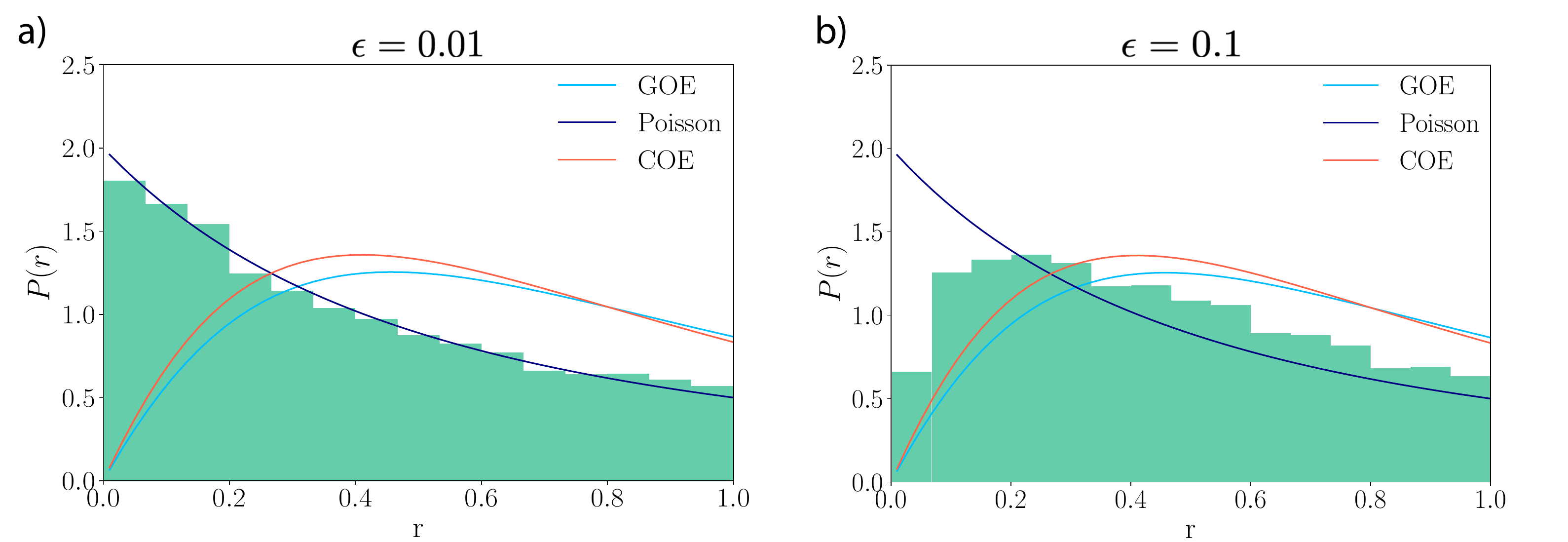}
	\caption{
		Statistical analysis of the ratios $r_{\alpha}=\min(\delta_{s},\delta_{s+1})/\max(\delta_{s},\delta_{s+1})$ between quasienergy gaps $\delta_{s}=\lambda_{s+1}-\lambda_{s}$. To obtain enough data, we calculate the quasienergy spectrum of a spin chain with $n=8$ sites for $50$ realizations of disorder. a) For a small value $\epsilon=0.01$ of the rotation error, the levels are uncorrelated and the statistical behavior resembles a Poissonian distribution, which is a signature of MBL phase. b) By increasing the error to $\epsilon=0.1$, the statistics is closer to the circular orthogonal ensemble. As long as the system remains in the MBL phase, the time-crystalline order can still be obserevd. We considered the same parameters used in the main text.
			} 
	\label{fig1}
\end{figure}

One of the most intriguing aspects of level statistics is its intimate relation to quantum signatures of chaos and random matrix theory. At this stage, it is important to clarify a technical point concerning the transition between regular and chaotic motion and its intimate relation to the MBL-to-ergodic transition in manybody systems. As we discussed above, the MBL phase is characterized by an extensive number of conserved quantities, which leads to a Poissonian level statistics. Under the effect of a perturbation, these conserved quantities are destroyed and the system has a transition to the ergodic phase. This process resembles the destruction of tori in phase space within the framework of KAM theorem. For small perturbations, some tori are more susceptible to be destroyed and as we increase the strength of the perturbation, the phase space becomes mixed: there are regular islands associated to the surviving tori and a chaotic sea. When all the tori are destroyed, the dynamics is fully chaotic. In quantum systems, signatures of this chaotic dynamics lead to a universal behavior of the level statistics giving rise to distributions $P_{\text{GOE}}(r)$ and $ P_{\text{COE}}(r)$ for undriven and driven systems, respectively. Universality means that these distributions do not depend on microscopic details of the dynamics but are solely determined by symmetries of the system. However, quantum signatures of a mixed phase space are not universal. In the context of time crystals, when we increase the error, we see a crossover between $P_{\text{Poisson}}(r)$ and $P_{\text{COE}}(r)$, because the error destroys conserved quantities. As we increase the rotation error, we observe that the level statistics is closer to $P_{\text{COE}}(r)$.   Figure~\ref{fig1} depicts the results for the level statistics in the case of a time crystal for errors $\epsilon=0.01$ and $\epsilon=0.1$. Our results therefore show that time crystalline order is protected by MBL.

\section{Power spectrum of time series: Stability of the time-crystalline phase}
\addcontentsline{toc}{subsubsection}{III. Power spectrum of time series: Stability of the time-crystalline phase}
\label{SI3}
In this section our aim is to provide an additional experimentally-accessible way to investigate the stability of the different configurations. The latter will shed light on the melting mechanism of the time crystal and the emergent scale-free behavior. Our idea is based on a quantity that can be easily measured in several platforms. We just require to measure the expectation value of the total magnetization $M_{i,Z}(mT)=1/n\sum_{r=1}^n\langle\psi(mT)|\sigma_r^z|\psi(mT)\rangle$  at stroboscopic times, where $|\psi(0)\rangle=|i\rangle$. This quantity is related to the populations, as follows
\begin{equation}
 \label{eq:TransverseMagPopulation}
M_{i,Z}(mT)=-\frac{1}{n}\sum_{r=1}^n\sum_{j=1}^{2^n}(-1)^{j_r}|A_{j} (mT) |^2
\ ,
\end{equation}
where 
 $(j_1,j_2,\ldots,j_n)_2$ is the binary decomposition of the integer number $j$ for all the configurations. 
 
 \begin{figure}[h!]
 	\centering
 	\includegraphics[width=0.75\textwidth]{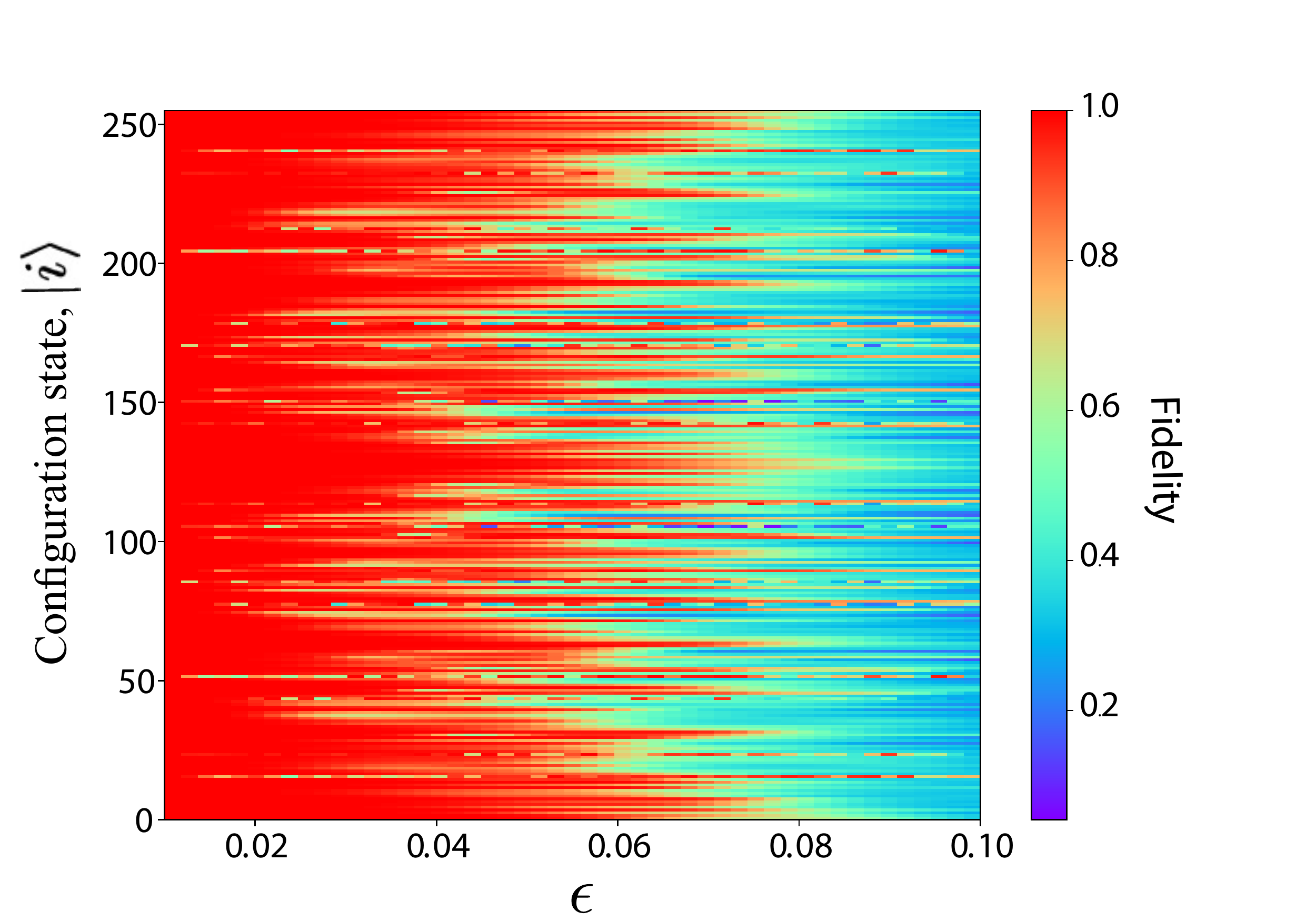}
 	\caption{
 		Stability of the different $2^8$ configuration states. To investigate the effect of the rotation error on the different configurations $|i\rangle$, we calculate the fidelity $F_{i}^{\text{ps}}(\epsilon)$. The latter is a measure of the periodicity of the dynamics. Our results show that configurations such as $|2^{n}-1\rangle=|\uparrow\uparrow\uparrow\ldots\uparrow\uparrow\rangle$ are more stable under the effect of error. Others, on the contrary, are strongly affected by the rotation error.  We consider a time crystal with $n=8$ sites ($256$ configurations) and the same parameters used in the main text.
 	} 
 	\label{fig3}
 \end{figure}
 
 Next, for a given initial configuration $i$, one can record the measurements $M_{i,Z}(mT)$, thus obtaining a time series 
 \begin{equation}
 \label{eq:TimeSeries}
 \{M_{i,Z}(0),M_{i,Z}(T),\dots,M_{i,Z}(NT)\}
 \ ,
 \end{equation}
 where $N$ is the number of periods that the system evolve. With the data for the time series at hand, we can calculate the discrete Fourier transformation
\begin{equation}
 \label{eq:FourierTransverseMagPopulation}
\mathcal{M}_{i,Z}(k)=\frac{1}{N}\sum_{m=1}^Ne^{-\frac{\mathrm{i}2\pi k}{N}m}M_{i,Z}(mT)=\frac{1}{N}\sum_{m=1}^Ne^{-\mathrm{i}\omega_k m T}M_{i,Z}(mT)
\ ,
\end{equation}
where $\omega_k=2\pi k/NT$ and  $k\in[0,N-1]$. For a given rotation error $\epsilon$, the associated power spectrum 
\begin{equation}
 \label{eq:FourierPowerSpec}
\boldsymbol{V}_{i,\epsilon}=\{|\mathcal{M}_{i,Z}(0)|^2,|\mathcal{M}_{i,Z}(1)|^2\dots,|\mathcal{M}_{i,Z}(N-1)|^2\}
\end{equation}
 tell us how strong is the contribution of a the $k-$th harmonic to the time series. Note that for convenience, we have arranged the Fourier coefficients in the form of a vector $\boldsymbol{V}_{i,\epsilon}$. In the absence of error, all the configurations show a magnetization that varies as $M_{i,Z}(mT)=(-1)^mM_{i,Z}(0)=e^{\mathrm{i}n\pi}M_{i,Z}(0)
$ and the power spectrum shows a single peak at a frequency $\omega_{N/2}$, which is half the frequency of the drive. The latter is a signature of the subharmonic response of the manybody system. As we discussed in Sec.\ref{SI2}, the periodic motion is related to local integrals of motion. The DTC is protected agains heating due to MBL. However, as we increase the rotation error $\epsilon$, some configurations are more affected than others and they loose their periodic character. On the contrary, configurations like $|2^{n}-1\rangle=|\uparrow\uparrow\uparrow\ldots\uparrow\uparrow\rangle$ and $|0\rangle=|\downarrow\downarrow\downarrow\ldots\downarrow\downarrow\rangle$ are the most stable ones. To have a quantitative measurement of this statement that can be easily accessed in experiments, we propose to measure the fidelity $F_i^{\text{ps}}(\epsilon)$ of the power spectrum, defined as follows
\begin{equation}
 \label{eq:Fidelity}
F_{i}^{\text{ps}}(\epsilon)=\sqrt{\frac{\boldsymbol{V}_{i,0}\cdot\boldsymbol{V}_{i,\epsilon}}{\|\boldsymbol{V}_{i,0}\|\:\|\boldsymbol{V}_{i,\epsilon}\|}}
\ .
\end{equation}
This very simple quantity tell us how far is the power spectrum from the one of a time crystal without error. In figure~\ref{fig3} we depict the fidelity as a function of the configuration number $i$ and the error $\epsilon$. This shows a clear evidence that not all the configurations are stable under the effect of a rotation error.

\section{Semiclassical description of a discrete time crystal: a stability analysis of the fixed points in the absence of error $\epsilon=0$}
\addcontentsline{toc}{subsubsection}{IV. Semiclassical description of a discrete time crystal: a stability analysis of the fixed points in the absence of error $\epsilon=0$}
\label{SI4}
In this section, we present numerical evidence of the stability of the different configurations from a semiclassical perspective. We start by considering the effective Hamiltonian for two periods of the drive

\begin{align}
      \label{eq:EffectiveHamiltonianNonZeroErrorS}
             \hat{H}_{\epsilon,2T}^{\text{eff}}&=\frac{\hbar T_2}{T}\sum_{lm}J_{lm}^{z}\sigma_{l}^{z} \sigma_{m}^{z}-\frac{\hbar g\epsilon T_1}{2T}\sum_{l}[(\cos{(B_l2T_2)}+1) \sigma_{l}^{x}  +\sin{(B_l2T_2)} \sigma_{l}^{y}] \ ,
\end{align}
which describes the dynamics of the system at stroboscopic times $t_n=2nT$.

After visualizing the classical orbits on the Bloch sphere, it is instructive to investigate a limiting case of the system. Here we focus on the system in the absence of error. In this case, there is an exact mapping of Eq.\eqref{eq:EffectiveHamiltonianNonZeroError} to classical Hamiltonian

\begin{equation}
\label{eq:ClassicalHam}
       \mathcal{H}_{0,2T}(\theta_1,\ldots,\theta_n)=\frac{\hbar T_2}{T}\sum_{l,m}J^z_{l,m}\cos\theta_l\cos\theta_m
       \ .
\end{equation}
Configurations such as $\theta^c_l=0,\pi$ are fixed points of the stroboscopic dynamics. The latter can be derived from the first derivative of the energy function $\partial_{\theta_j}\mathcal{H}_{0,2T}(\theta_1,\ldots,\theta_n)=0$.
For example, the quantum  $|2^n-1\rangle=|\uparrow\uparrow\uparrow\ldots\uparrow\uparrow\rangle$ has a classical counterpart $(\theta^c_1,\theta^c_2,\ldots,\theta^c_n)=(0,0,\ldots,0)$.  Given a fixed point, we can explore its stability properties by investigating the eigenvalues of the Jacobian matrix with elements $\mathcal{J}_{i,j}=\partial_{\theta_i}\partial_{\theta_j}\mathcal{H}_{0,2T}(\theta_1,\ldots,\theta_n)|_{\theta=\theta^c}$, where the derivatives are evaluated at the critical points $\theta^c$ such that $\partial_{\theta_j}\mathcal{H}_{0,2T}(\theta_1,\ldots,\theta_n)=0$. For simplicity, let us focus on a particular example of a lattice with four sites and open boundary conditions. In the case of the configuration $(\theta^c_1,\theta^c_2,\theta^c_3,\theta^c_4)=(0,0,0,0)$
these matrix elements are given by
\begin{align}
\label{eq:JacMatrix}
       \mathcal{J}_{i,i}&=-\sum_{m}J^z_{i,m}\cos\theta_i\cos\theta_m|_{\theta=\theta^c}=-\sum_{m}J^z_{i,m}
       \nonumber\\
       \mathcal{J}_{i,j}&=J^z_{i,j}\sin\theta_i\sin\theta_j|_{\theta=\theta^c}=0
       \ .
\end{align}
and the equilibrium is stable \cite{Engelhardt2013}. For any configuration of the form $\theta^c_l=0,\pi$ the off-diagonal elements of the Jacobian matrix are zero.
Let us consider the configuration $(\theta^c_1,\theta^c_2,\theta^c_3,\theta^c_4)=(0,0,\pi,\pi)$ with one domain wall, where $ \mathcal{J}_{1,1}=-J^z_{1,2}+J^z_{1,3}+J^z_{1,4}$, $ \mathcal{J}_{2,2}=-J^z_{1,2}+J^z_{2,3}+J^z_{2,4}$, $ \mathcal{J}_{3,3}=-J^z_{1,3}+J^z_{2,3}-J^z_{3,4}$ and $ \mathcal{J}_{4,4}=J^z_{1,4}+J^z_{2,4}-J^z_{3,4}$. Due to the power-law decay of the couplings, all the diagonal elements are negative and the fixed point is stable. Finally, for a configuration with three domain walls $(\theta^c_1,\theta^c_2,\theta^c_3,\theta^c_4)=(0,\pi,0,\pi)$, the matrix elements are  $ \mathcal{J}_{1,1}=J^z_{1,2}-J^z_{1,3}+J^z_{1,4}$, $ \mathcal{J}_{2,2}=J^z_{1,2}+J^z_{2,3}-J^z_{2,4}$, $ \mathcal{J}_{3,3}=-J^z_{1,3}+J^z_{2,3}+J^z_{3,4}$ and $ \mathcal{J}_{4,4}=J^z_{1,4}-J^z_{2,4}+J^z_{3,4}$. In this case all the eigenvalues are positive. The elements of the Jacobian give us information about the local curvature of the energy function at the critical points. As long as they have the same sign, the fixed points are stable. For a non-zero value of the error, configurations such as $\theta^c_l=0,\pi$ are not fixed points of the dynamics anymore, but they belong to trajectories in phase space. 

\section{Quantum walks in the configurations space: Experimental protocol to simulate complex quantum networks}
\addcontentsline{toc}{subsubsection}{V. Quantum walks in the configurations space: Experimental protocol to simulate complex quantum networks}
\label{SI5}
So far, we have discussed spectral properties of driven systems and its consequences for the time crystal. The percolation rule discussed in the main text provides us with graphical way to represent localization properties of Floquet states. Therefore, the connectivity of the graph tell us information about how many configurations ``participate" in a given Floquet state. When the latter are delocalized, the gaps between quasienergies become highly correlated and the system becomes ergodic as we discussed in section \ref{SI2}. This establish an intriguing relation between the melting of a discrete time crystal and the MBL-to-ergodic transition as we increase the rotation error. During this process, the graph exhibits a scale-free behavior characteristic of complex quantum networks. Here, we describe in detail an experimental protocol to faithfully extract structural information about the simulated complex quantum networks in Noisy intermediate-scale quantum devices. 

The cornerstone of our approach is the  representation of the effective Hamiltonian
\begin{equation}
\label{eq:TightBinding}
\hat{H}^{\text{eff}}_{\epsilon,T}=\sum_{i} \mathcal{E}_{i}|i\rangle \langle i| + \sum_{i,j} K_{ij}|i\rangle \langle j|=\sum_{s} \lambda_{s}|\Phi_s\rangle \langle \Phi_s|
\end{equation}
in terms of the configuration basis states $\lbrace |i\rangle \rbrace$. Here $\mathcal{E}_{i}$ is the energy of configuration $|i\rangle$, and $K_{ij}$ is the transition energy between configurations $|i\rangle$ and $|j\rangle$. The quasienergies $\lambda_{s}$ have a complicated dependence on the parameters $\mathcal{E}_{i}$ and $K_{ij}$ that determine localization properties of the Floquet states $|\Phi_s\rangle$. The latter are linear combinations of configurations $|i\rangle$ with $i=0,1,2\ldots,2^{n}-1$, where $n$ is the number of sites of the spin chain.
Interestingly, although the effective Hamiltonian describes a manybody system, in terms of configurations, the dynamics resemble a quantum walk in a quantum complex network. 

\begin{figure}[h!]
	\centering
	\includegraphics[width=0.95\textwidth]{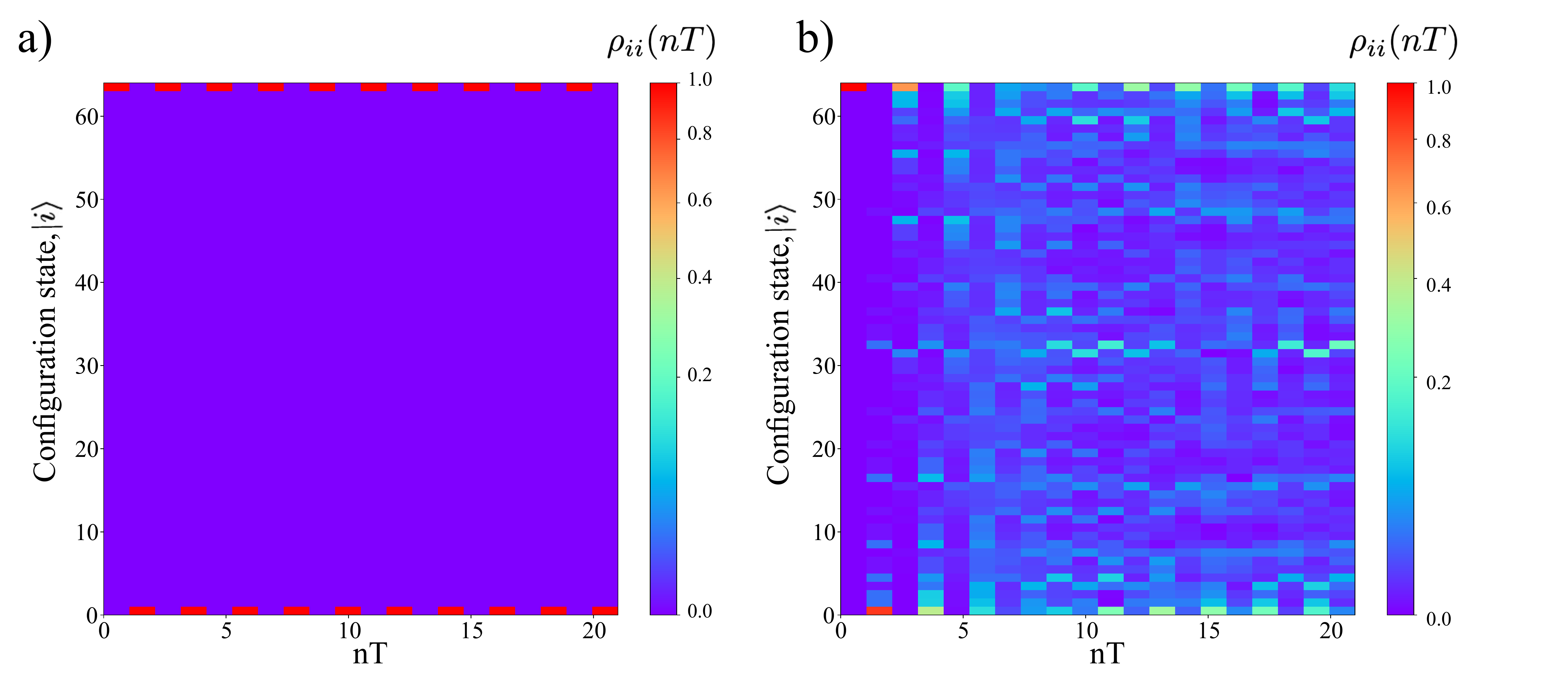}
	\caption{
		Quantum simulation of a complex quantum network of $n=6$ sites. One of the cornerstone of our work is the possibility to use time crystals as a quantum simulator for a complex quantum network. In terms of the configurations $|i\rangle$, the effective Hamiltonian~\ref{eq:TightBinding} is a tight binding model on a complex quantum network. To experimentally simulate the dynamics of an excitation in such a network, one can prepare a configuration $|\psi(0)\rangle=|i\rangle$ as the initial state of the system. For example, we can prepare $|\psi(0)\rangle=|2^{n}-1\rangle=|\uparrow\uparrow\uparrow\ldots\uparrow\uparrow\rangle$ and register the stroboscopic evolution of the populations $\rho_{i,i}(nT)$ of the different configurations $|i\rangle$ at stroboscopic times $t_n=nT$. a)  Depicts the evolution of the populations in the absence of error $\epsilon=0$. In this case, the network is composed by a set of decoupled dimers. Therefore, the dynamics takes place between two configurations  $|0\rangle$ and $|2^{n}-1\rangle$. b) Shows the dynamics of the populations for an error $\epsilon=0.1$. For this value of the error, the complex network exhibits clusters and the system can populate more configurations in the Hilbert space.
		We consider a time crystal with $n=6$ sites ($64$ configurations) and the same parameters used in the main text.
	} 
	\label{fig2}
\end{figure}

Experimentally, one can initially prepare the state of the system in a given configuration such as $|\psi(0)\rangle=|i\rangle=|\uparrow\uparrow\uparrow\ldots\downarrow\downarrow\rangle$, which represents a node of the complex network. At stroboscopic times $t_n=nT$, the dynamics are governed by the effective Hamiltonian $\hat{H}^{\text{eff}}_{\epsilon,T}$ and the evolution of the state reads $|\psi(nT)\rangle=e^{-\mathrm{i}\hat{H}^{\text{eff}}_{\epsilon,T}nT/\hbar}|\psi(0)$. Inherently, the most relevant aspects of the dynamics can be recovered from this discrete evolution. Next, let us focus on the density matrix 
\begin{equation}
\label{eq:DensityMatrix}
\hat{\rho}(nT)=|\psi(nT)\rangle\langle\psi(nT)|= \sum_{i,j}A_{i} (nT) A^{*}_{j}(nT)|i\rangle \langle j|,
\end{equation}
where $|\psi(nT)\rangle=\sum_{i}A_{i} (nT)\|i\rangle $. This expression give us already a recipe to measure the dynamics of the quantum walk in existing Noisy intermediate-scale quantum devices. This requires a modest set of measurements and it is straightforward with current experimental feasibilities.  The idea is to measure the populations $\rho_{i,i}(nT)=|A_{i} (nT) |^2$ of the density matrix stroboscopically, to determine which configurations are occupied. This information can be accessed by measuring the correlation functions

\begin{align}
\label{eq:Generalcorrelation}
C_{i,Z}(nT)&=\text{tr}\left[\hat{\rho}(nT)|i\rangle \langle i|\right]=|A_{i} (nT) |^2
\nonumber\\ &=
\frac{1}{2^n}\text{tr}\left[\hat{\rho}(nT)(\sigma_1^z-(-1)^{i_1}\hat{I})(\sigma_2^z-(-1)^{i_2}\hat{I})\cdots(\sigma^{z}_n-(-1)^{i_n}\hat{I})\right]
\ ,
\end{align}
corresponding to all the configurations $i=0,1,2,\dots,2^n-1$. Here
$(i_1,i_2,\ldots,i_n)_2$ is the binary decomposition of the integer number $i$. 
For example, if we want to measure the population of the configuration $|2^n-1\rangle=|\uparrow\uparrow\uparrow\ldots\uparrow\uparrow\rangle$, we need to measure the correlation
\begin{equation}
\label{eq:CorrEx1}
C_{2^{n-1},Z}(nT)=\frac{1}{2^n}\text{tr}\left[\hat{\rho}(nT)(\sigma_1^z+\hat{I})(\sigma_2^z+\hat{I})\cdots(\sigma^{z}_n+\hat{I})\right]
\ ,\end{equation}
because $2^{n-1}=(1,1,\ldots,1)_2$.
Similarly, to measure the population of $|0\rangle=|\downarrow\downarrow\downarrow\ldots\downarrow\downarrow\rangle$, we need to measure the correlation
\begin{equation}
\label{eq:CorrEx2}
C_{0,Z}(nT)=\frac{1}{2^n}\text{tr}\left[\hat{\rho}(nT)(\sigma_1^z-\hat{I})(\sigma_2^z-\hat{I})\cdots(\sigma^{z}_n-\hat{I})\right]
\ ,
\end{equation}
as $0=(0,0,\ldots,0)_2$. Interestingly, the aforementioned correlations can be reconstructed from measurements of all the possible combinations of single-point correlators $C_{Z}^{r_1}(nT)=\text{tr}\left[\hat{\rho}(nT)\sigma_{r_1}^z\right]$, two-point correlators $C_{Z}^{r_1,r_2}(nT)=\text{tr}\left[\hat{\rho}(nT)\sigma_{r_1}^z\sigma_{r_2}^z\right]$ up to n-point correlators $C_{Z}^{r_1,r_2,\ldots,r_n}(nT)=\text{tr}\left[\hat{\rho}(nT)\sigma_{r_1}^z\sigma_{r_2}^z\cdots\sigma_{r_n}^z\right]$. In trapped ions, it is possible to measure these correlations by resorting to measurements on single ions. In superconducting qubits, it is even possible to perform full tomography for qubits arrays up to $N=12$ qubits and measurements of these correlations are within reach with current technology~\cite{Gong2019}. After recording the data of the populations $|A_{i} (nT) |^2$ for all the configurations $i$ at stroboscopic times $t_n=nT$, one can generate a space-time plot to depict the data $\left[nT, i,\rho_{i,i}(nT)\right]$. The latter is depicted in Fig.~\ref{fig2}. The color scale represent the populations and one can visualize how the initial states travel through the configuration space.

The experimental protocol mentioned above provides us with a way of accessing very relevant information related to its degree distribution. To have a simple picture of this idea in mind, let us consider the case of a DTC in the absence of rotation error. In this case, if we prepare an initial configuration $|\psi(0)\rangle=|\uparrow\uparrow\uparrow\ldots\downarrow\downarrow\rangle$, this correspond to a Floquet cat state, because it is a quantum superposition of two Floquet states with quasienergies $\pm \hbar\pi/T$. Therefore, the expectation value of the total transverse magnetization $M_{i,Z}(nT)=1/n\sum_{r=1}^n\langle\psi(nT)|\sigma_r^z|\psi(nT)\rangle$ will show an oscillatory behavior with a period $2T$. In the configuration space, the quantum walk will show dynamics between just two configurations $|2^{n}-1\rangle=|\uparrow\uparrow\uparrow\ldots\uparrow\uparrow\rangle$ and $|0\rangle=|\downarrow\downarrow\downarrow\ldots\downarrow\downarrow\rangle$. If we now calculate how many configurations "participate" in the quantum state $|\psi(nT)\rangle$, then we have a measure of how connected the network in the configuration space is. To be more precise, let us define the participation ratio
\begin{equation}
\label{eq:ParticipationRatio}
P_i(nT)=\frac{1}{\sum_{i=1}^{2^n}|A_{i} (nT) |^4}
\ ,
\end{equation}
which can be obtained from the measurement of the correlation function Eq.~\eqref{eq:Generalcorrelation}. 

\begin{figure}[h!]
	\centering
	\includegraphics[width=0.8\textwidth]{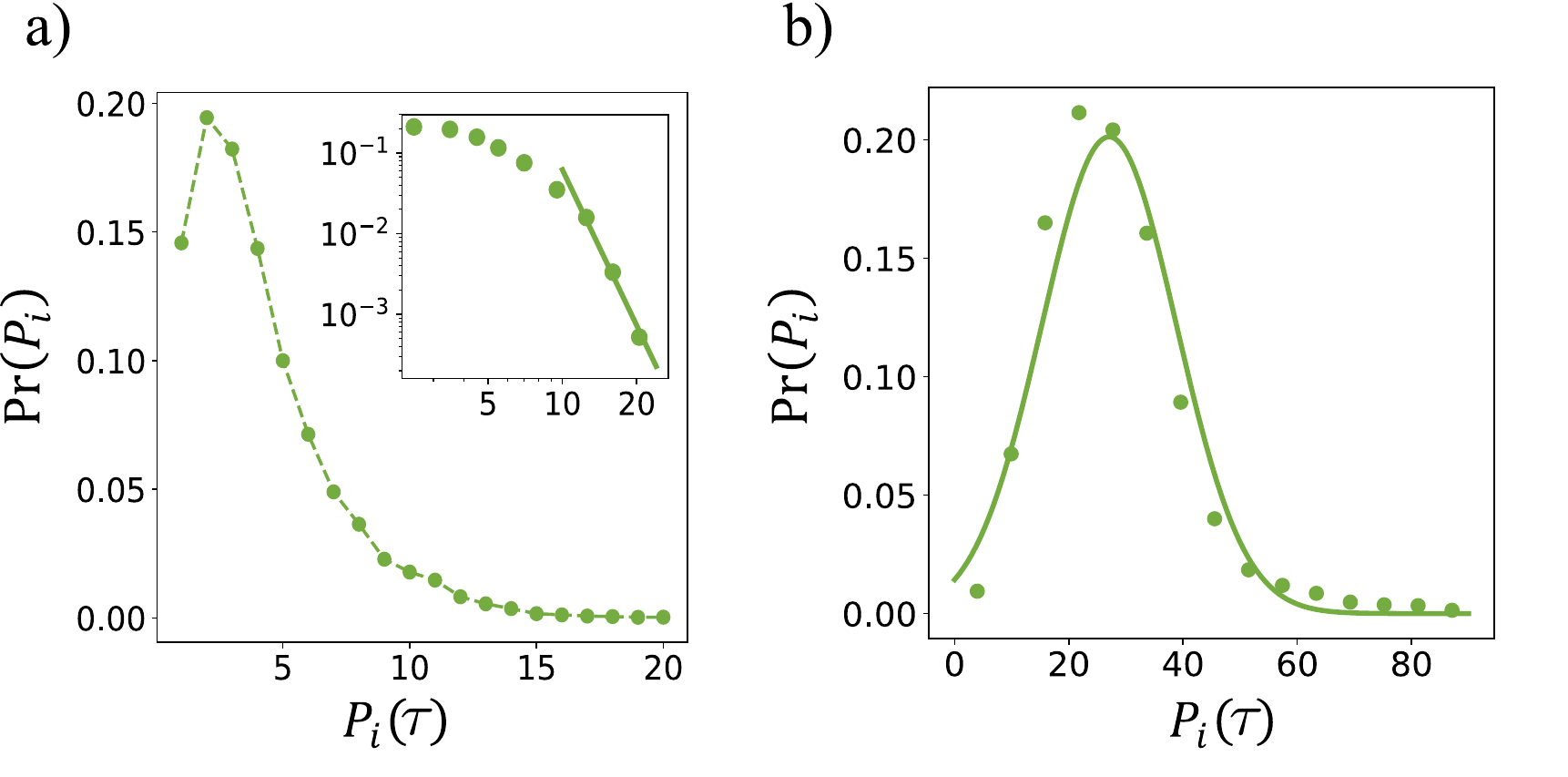}
	\caption{
		$PT$ distributions of a quantum network of $n=8$ sites as an approximation of the degree distribution of the network using only experimentally accessible data. Information regarding the complexity of the network can be gathered experimentally by the measurement of the correlation function of Eq.~\ref{eq:CorrEx1} and the calculation of the participation ratio, $P_i(\tau)$, at time $\tau\sim\frac{T}{g\epsilon T_1}$ for all the possible initial configurations $i=0,1,2,\ldots,2^{n}-1$. a) Depicts the $PT$ distribution for a $\epsilon=0.012$. The distribution is shown in both liner and logarithmic scale (see inset) and display a heavy-tailed distribution that can be fitted with a power-law curve. This distribution resembles very much to the one presented in Fig.~\ref{fig:TDTCgraph}.\textbf{c)}, where the presence of large degree hubs would indicate a scale-free feature of the complex network. b) Shows the $PT$ distribution for a $\epsilon=0.1$. This distribution has the shape of a normal distributions, similar to what was shown in Fig.~\ref{fig:TDTCgraph}.\textbf{f)} for the same value of $\epsilon$.
	} 
	\label{fig4}
\end{figure}

When approximating $\hat{H}^{\text{eff}}_{\epsilon,T}$ up to a first order in the error $\epsilon$, one can easily see that the local rotation that connects different resonant configurations (and therefore generates the clusters in the graph) has an approximate tunneling time of $\tau\sim\frac{T}{g\epsilon T_1}$. If we obtain $P_i(\tau)$ for a given initial $i$-th node or configuration, one can estimate how many configurations have been visited after the evolution time $\tau$, and thus estimate the degree of this node. If we now repeat the process for all the possible initial configurations $i=0,1,2,\ldots,2^{n}-1$ one can obtain a distribution for the participation ratio, $PR$. In Fig.~\ref{fig4} we present the distributions obtained from this experimental protocol for the same system that was depicted in Fig.~\ref{fig:TDTCgraph} (network of $n=8$ sites and a rotation error of $\epsilon=0.012$ and $\epsilon=0.1$). We note that the shape of the distributions obtained from the calculation of $P_i(\tau)$, \textbf{a)} heavy-tailed with a power-law fit and \textbf{b)} normal distribution, matches the ones of Fig.~\ref{fig:TDTCgraph} obtained directly from the degree distribution of the network. In a nutshell, by measuring the aforementioned correlation functions, we can simulate the network and obtain important information without calculating the unitary operator and without full quantum process tomography.

\end{document}